This paper is a preprint of a paper submitted to *IET Microwaves, Antennas and Propagation*. If accepted, the copy of record will be available at the IET Digital Library

# Theory and Design of a Phase-Inverted Balanced Coupled-Line DC-Blocker

Mostafa Abdolhamidi and Mahmoud Mohammad-Taheri

***Abstract*—A planar DC-blocker suitable for differential mode signaling applications is designed and fabricated. The theory of this component is explained in a new form which utilizes the wave scattering transfer matrix. The proposed interpretation of the transfer matrix is most suitable for series (cascade) elements like DC-blockers. In addition to the theoretical enhancement, design of a compressed balanced DC-blocker inserted through a shielded broadside coupled stripline (SBCSL) transmission line is presented. The return loss of better than 10 dB is obtained at 50-ohm differential-mode input ports of the fabricated DC-blocker in the entire frequency range of 5.6-8.4 GHz. The lowest air-gap width in the presented structure is about 10 times bigger than that of a conventional coupled-line structure. So, the structure is much less sensitive to fabrication tolerances. Moreover, the DC-blocker is likely to tolerate higher DC-voltage differences. Also, a demonstration for a millimeter-wave version of this DC-blocker suitable for integrated circuits (ICs) applications is proposed for future development. The final achievement of this paper is design and fabrication of a wideband substrate integrated waveguide (SIW)-mediated balun structure for single-ended measurement of a balanced SBCSL component. The fabricated balun exhibits a nearly perfect coaxial-mode to coupled-stripline differential-mode conversion in the full range of 5-9 GHz. The presented balun is successfully utilized to derive the scattering parameters (S-parameters) of the fabricated balanced SBCSL DC-blocker.**

## I. Introduction

THEORY of planar multi-conductor transmission lines (TLs) and coupled-line circuit elements are well established for the decades [1]-[3]. So far, various realizations for filters, directional couplers, baluns, etc. based on coupled-line elements have been proposed [4]. The series DC-blocking is a task well suited for coupled-line structures since the structure is wideband due to the intrinsic travelling wave effect of coupled lines. [4]. Beside the coupled-line realization, there exist other DC-blocking solutions: like discrete lumped capacitors for printed circuit boards (PCBs) and metal finger capacitors or metal-insulator-metal (MIM) capacitors for ICs [5]-[7]. At low frequencies, the lumped capacitors are used due to their small size. However, at higher frequencies the coupled-line DC-blockers exhibit lower losses, thus they are preferable [4]. In addition, for applications like transformer coupling of cascaded IC circuit stages [8]–[9], one can modify coupled-line DC-blockers in order to provide extra functionalities such as phase inversion, impedance matching and so on. Successful demonstration of transformer coupled active ICs for millimeter-wave applications can be found in the recent papers [10]-[13]. In this sense, a modified phase-inverted DC-blocker can be approximately seen as a unit-ratio transformer which can be utilized as an interface of two cascaded active stages with the same characteristic impedances (Fig. 1).

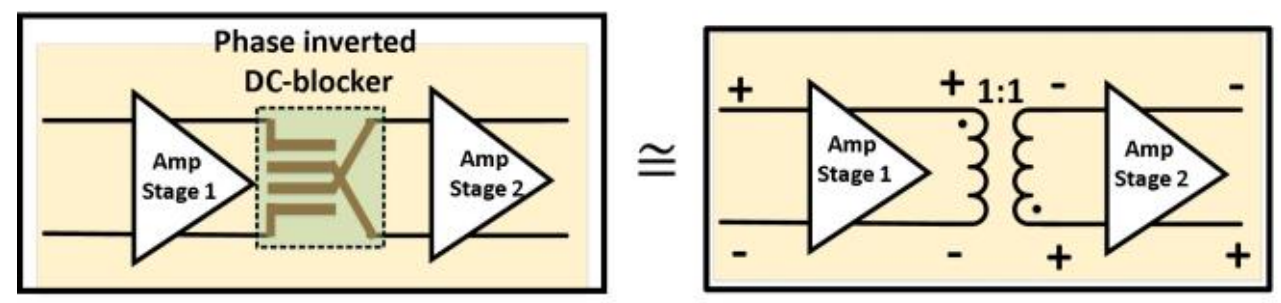

Fig. 1. Equivalence of a phase inverted DC-blocker to a 1:1 transformer.

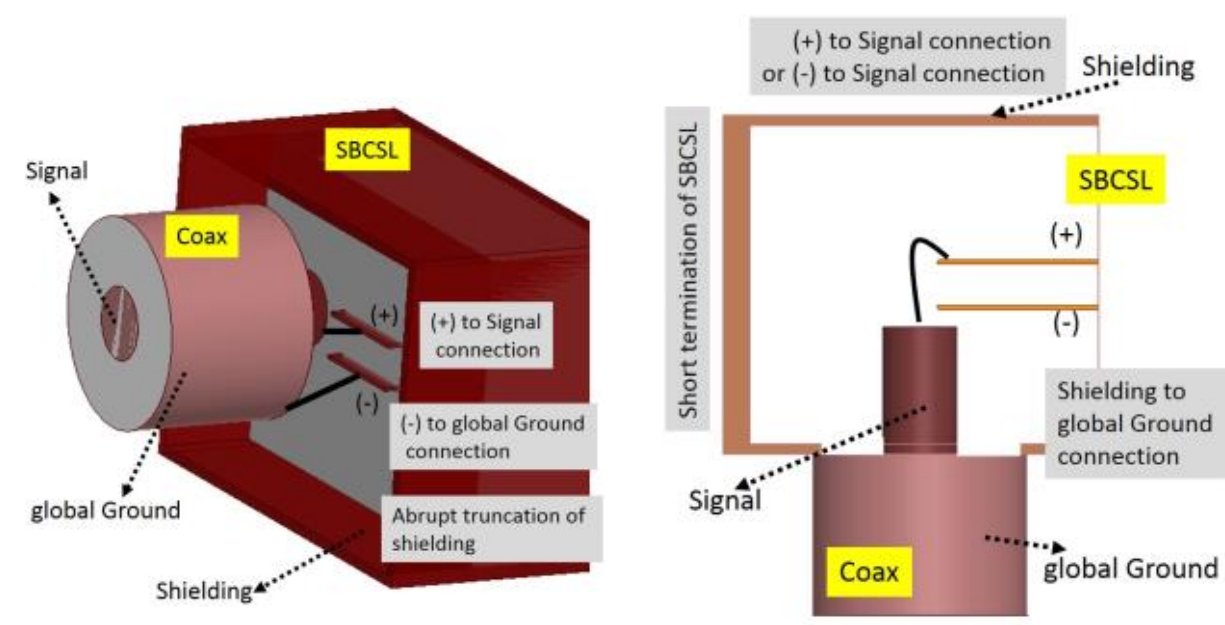

(a) (b)

Fig. 2. Difficulties in a double-conductor to triple-conductor balun transition design: a) abrupt truncation of the shielding b) electrical connection of the shielding to the global ground.

Previously, theory of design and operation of conventional coupled-line DC-blockers have been explained in terms of even and odd modes characteristic impedances [2], [4]. In this paper, using the TL capacitance and inductance matrices, we present an alternative derivation based on the wave scattering transfer matrices. This derivation is mostly fit for cascaded circuits where the wave scattering transfer matrix of the whole circuit is simply obtained by successive multiplication of wave scattering transfer matrices of all stages. The derivation of overall wave scattering transfer matrix of a multi-step DC-blocker is covered in section II. In addition, by the presented mathematical analysis, we show that the operational bandwidth of the DC-blocking stage will automatically

This work is under the support of Iran National Science Foundation (INSF).

Mostafa Abdolhamidi is with School of ECE, University of Tehran, Tehran, Iran. (email: abdolhamidi@ut.ac.ir)

Mahmoud Mohammad-Taheri is with School of ECE, University of Tehran, Tehran, Iran. (email: mtaheri@ut.ac.ir)

broaden by implementing the DC-blocking function inside a balanced TL. In the next section, we present a phase-inverted DC-blocker for SBCSL structure. Since aspect ratios (AR) of the metallic tracks in a PCB are normally very low [14], obtaining high values of couplings between adjacent metallic traces in a PCB is a difficult task and generally results in extremely close coupled traces. Our proposed design is a single-layer PCB structure in 5-10 GHz frequency range in which the loose coupling problem is properly solved by use of auxiliary couplings between metallic traces.

The last remarkable contribution of this work is the design of a novel coaxial line-to-SBCSL wideband balun which we use as an interface for the measurement of the DC-blocker S-parameters. In our proposed balun, we have used a SIW-based interface to solve the problems which usually appear in the connection of a shielded balanced TL to an unbalanced TL. These problems are namely the undesired radiation (single-frequency or wideband) (Fig. 2a), and the excitation of unwanted resonant modes within the balun structure (Fig. 2b). The former appears if the conductor shielding is truncated, while the latter is normally a consequence of short-end termination of the conductor shielding. In addition, the mentioned SIW interface facilitates the correction of balun phase and amplitude imbalance. This problem exists in balun transitions which have been designed based on lumped element [14]- [17] or coupled-line Marchand balun dividers [18]- [20]. Also, our proposed balun does not require high values of even to odd mode impedances ratio which is necessary in the solution proposed by [21]. The design procedure of the explained balun is also discussed in section III.

The fabrication and measurement processes are presented in section IV. In section V we show that the fabricated microwave PCB phase-inverted DC-blocker can be redesigned for millimeter wave IC applications. A simulation of a tightly compressed spiral-form DC-blocker in millimeter wave is presented as a future work in this section.

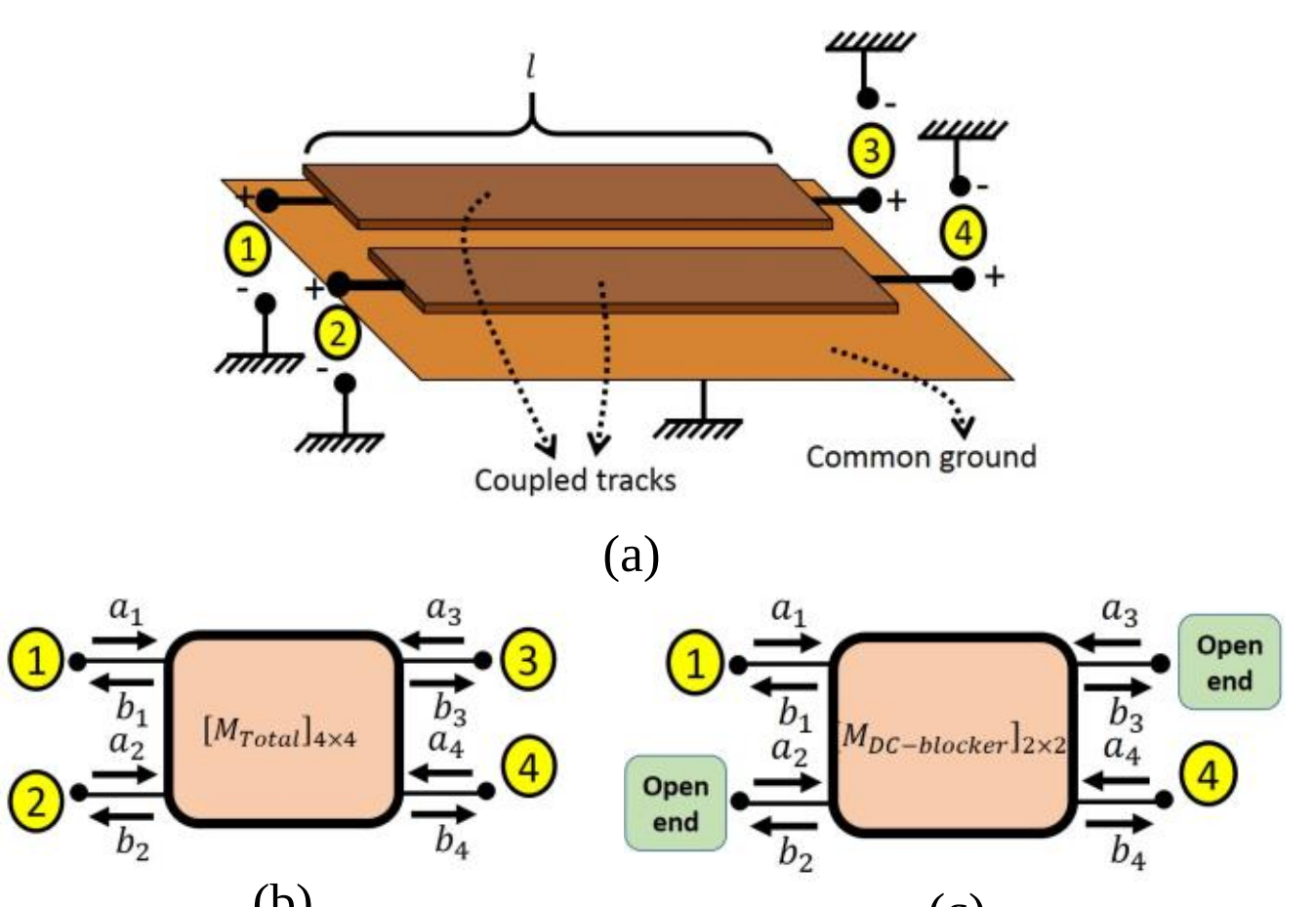

Fig. 3. Use of a pair of common-ground coupled tracks as a four-port network: a) arrangement of the coupled pair, b) its four-port equivalent network, c) two-port network which is to be used as the basis of the DC-isolated transition.

## II. Theory Improvement

### A. *Wave Scattering Transfer Matrix Representation*

Fig. 3a shows a pair of typical common-ground coupled traces. The depicted structure if is seen as a four-port network like that of Fig. 3b can be represented by a wave scattering transfer matrix ($M_{Total}$), using the procedure rigorously explained in Appendix. The input and output power waves are related to each other through $M_{Total}$ by:

$$\begin{pmatrix} a_1 \\ a_2 \\ b_1 \\ b_2 \end{pmatrix} = M_{Total} \begin{pmatrix} b_3 \\ b_4 \\ a_3 \\ a_4 \end{pmatrix} \quad (1)$$

According to the derivation given in Appendix, $M_{Total}$ is equivalent to an exponential matrix of the form:

$$M_{Total} = e^{j\theta[D^i]} \quad (2)$$

where $\theta$ is the coupled-line electrical length and $[D^i]$ is an involutory matrix (that means $[D^i] = [D^i]^{-1}$) defined by (A12) and (A15) in Appendix, respectively. Now, suppose we left ports (2) and (3) of Fig. 3b open to derive an expression for a wave scattering transfer matrix for a two-port network of Fig. 3c which is a simple circuit model for a one-stage DC-blocker. After some mathematical manipulation one can find that:

$$\begin{pmatrix} a_1 \\ b_1 \end{pmatrix} = M_{DC-blocker} \begin{pmatrix} b_4 \\ a_4 \end{pmatrix} \quad (3)$$

where

$$M_{DC-blocker} = \frac{-\frac{1}{2}}{M_{21}+M_{23}} \begin{pmatrix} M_{11}+M_{13} \\ M_{31}+M_{33} \end{pmatrix}_{2\times1} \begin{pmatrix} M_{22}-M_{42} & M_{24}-M_{44} \end{pmatrix}_{1\times2} - \begin{pmatrix} M_{12} & M_{14} \\ M_{32} & M_{34} \end{pmatrix}_{2\times2} \quad (4)$$

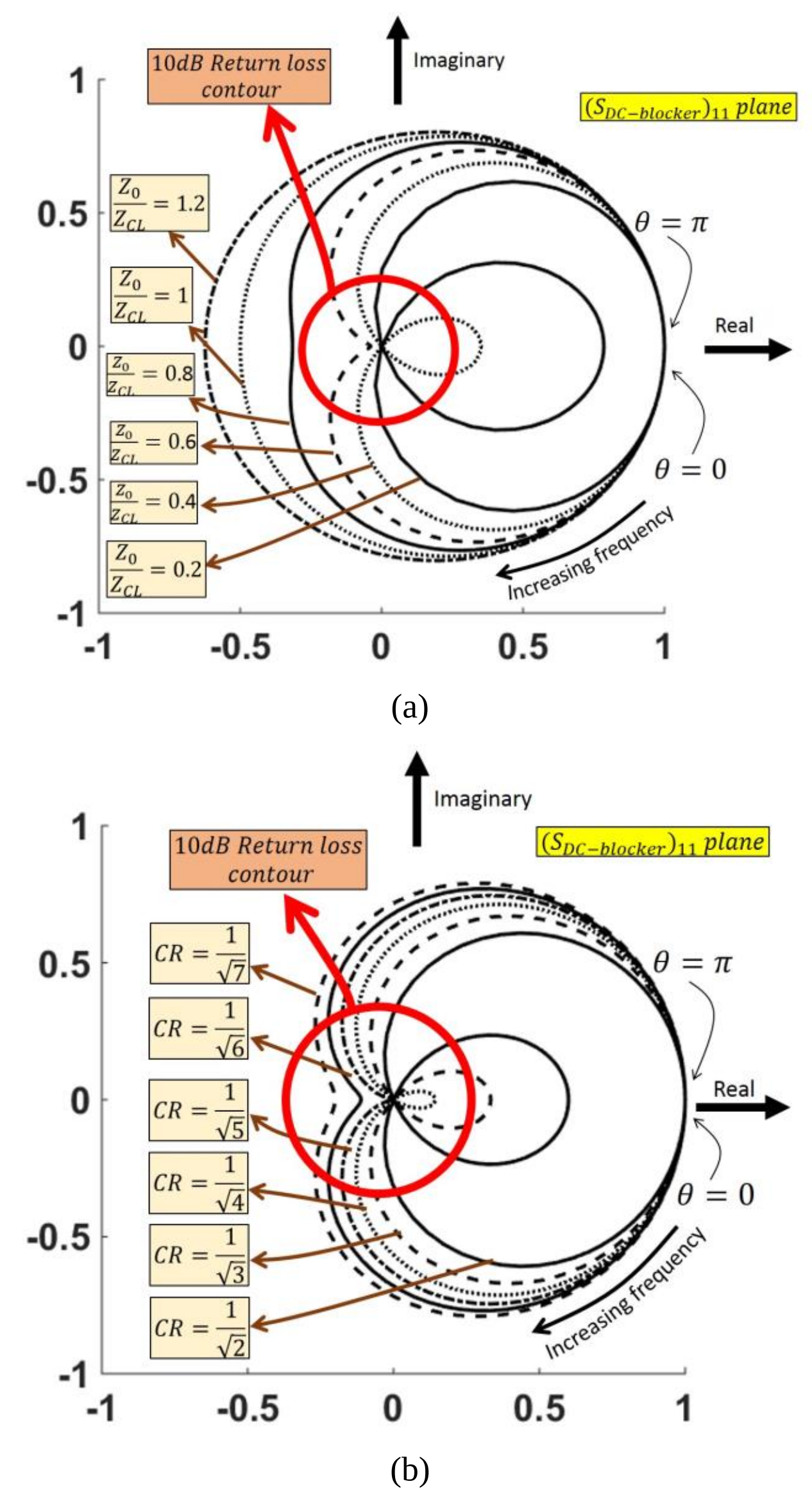

Fig. 4. Variations of $|(S_{DC-blocker})_{11}|$ as a function of frequency for different values of $\left(\frac{Z_0}{Z_{CL}}\right)$ and CR: a) $CR = 0.5$ contours, b) $\left(\frac{Z_0}{Z_{CL}}\right) = 0.5$ contours.

In (4), $(M_{ij})$s are the elements of the $(4\times4)$ matrix $M_{Total}$. According to (4), and (A12), (A15) and (A20) given in Appendix, $M_{DC-blocker}$ depends on the operating frequency ($\omega$), coupled-line section length ($l$), electromagnetic (EM) wave velocity ($v_{TE}$), the reference impedance ($Z_0$), and the capacitance per unit length matrix of the coupled-line structure ($C_u$). The advantage of using wave scattering transfer matrix as proposed above over conventional impedance matrix notations used in classic references [1], [3] is its compatibility with cascaded circuit architecture. For instance, overall $(4\times4)$ wave scattering transfer matrix for a multi-step (n-step) DC-blocker, is equal to:

$$M_{cascaded} = M_{Total}^{1} M_{Total}^{2} \ldots M_{Total}^{n} \quad (5)$$

or

$$M_{cascaded} = e^{j\theta_1[D_1^i]} \times e^{j\theta_2[D_2^i]} \ldots \times e^{j\theta_n[D_n^i]} \quad (6)$$

which can be simply written as

$$M_{cascaded} = \prod_{k=1}^{n} \left( cos\theta_k * I_{4\times4} + jsin\theta_k * [D_k^i] \right) \quad (7)$$

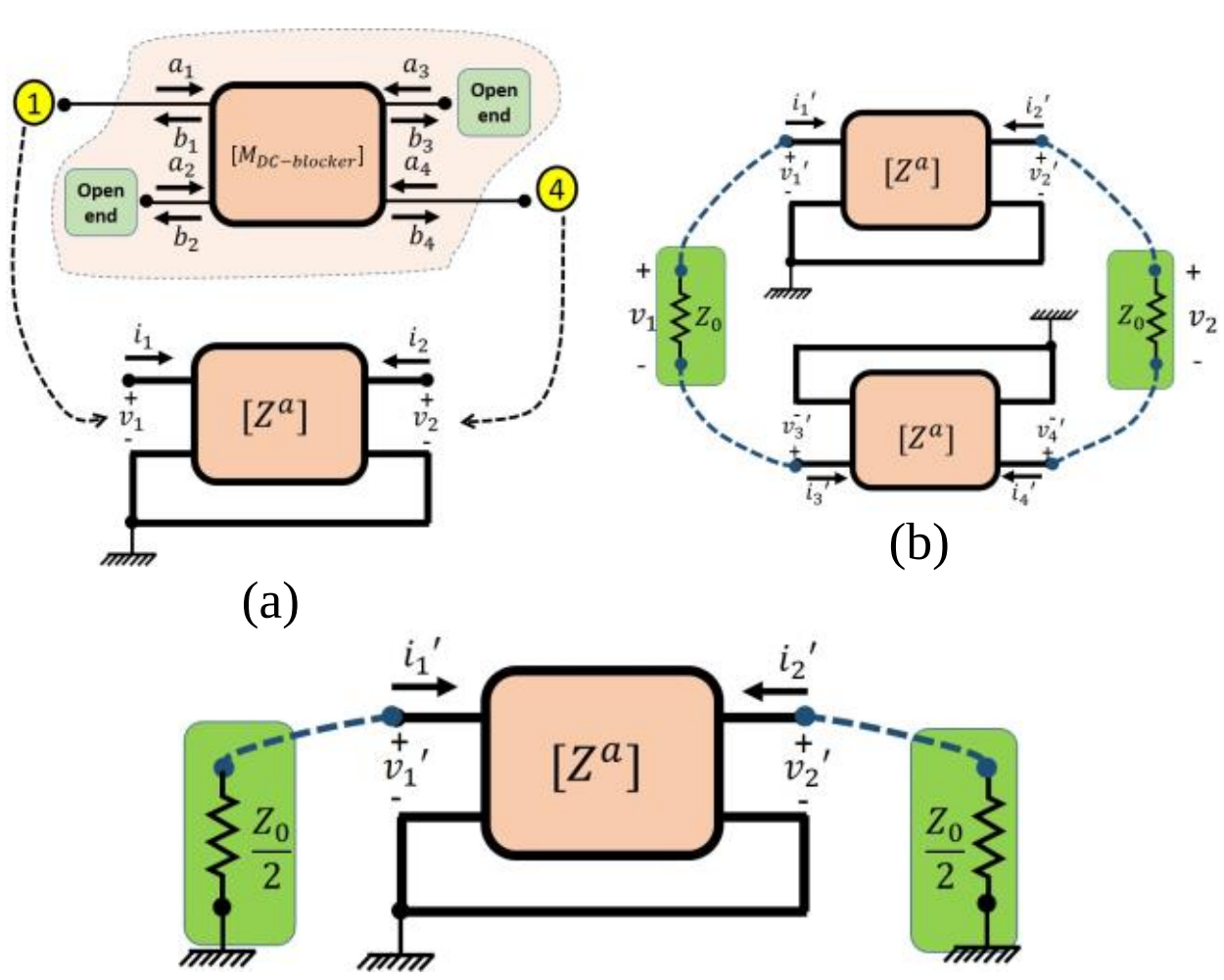

Fig. 5. Effect of differential mode excitation on the reference impedance of a two-port network: a) impedance representation of the structure of Fig. 3c, b) series connection of two networks with differential excitations, c) equivalent single-ended circuits of part (b).

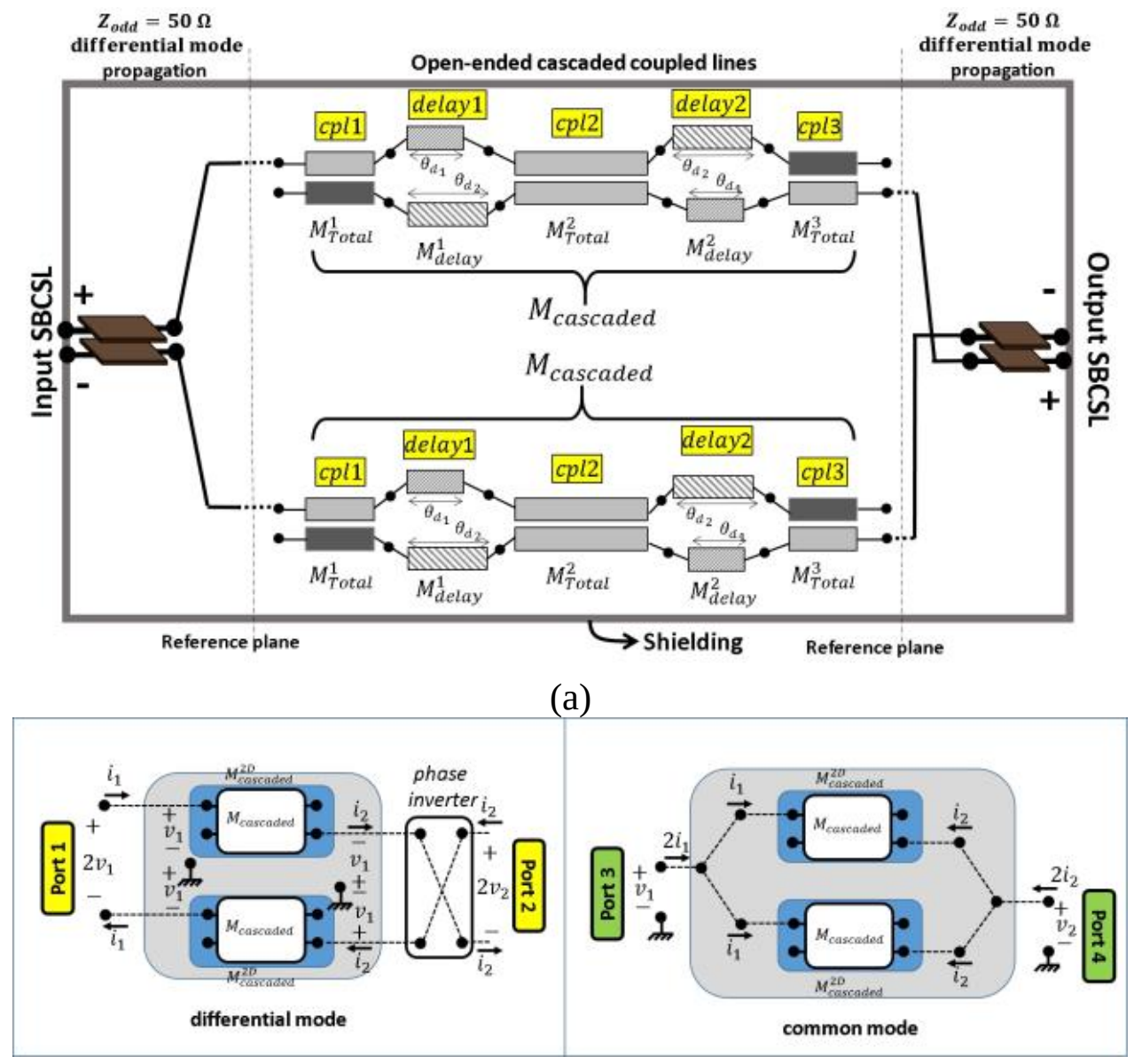

Fig. 6. Simple equivalent circuit of the proposed DC-blocker: a) circuit schematic diagram, b) differential mode and common mode excitations.

Then, the wave scattering transfer matrix of the n-step DC-blocker is simply calculated using (4). It is obvious that all other network parameters (like Z-parameters or S-parameters) can be directly calculated from the derived $M_{DC-blocker}$. No other network representation, namely the conventional Z-parameters representation has a simple form of (7) for cascaded circuit architecture when the port number is greater than 2.

## *B. Bandwidth Improvement in Balanced Architecture*

In this subsection it will be shown that for a balanced TL architecture, the performance of a coupled-line DC-blocker is better than that of other DC-isolation solutions. This is due to this fact that its frequency response is automatically improved when is used inside a balanced configuration. The scattering parameters of the two-port DC-blocker ($S_{DC-blocker}$) of Fig. 3c can be simply derived from $M_{DC-blocker}$. It is clear from (A20) in Appendix that, regardless of the configuration of the coupled-line cross section, the imaginary part of $\mathrm{M_{Total}}$ vanishes if

$$\theta = \frac{\omega l}{v_{TL}} = n\pi \quad (8)$$

This condition is identical to $(S_{DC-blocker})_{11} = 1$ and $(S_{DC-blocker})_{22} = 1$. Thus, our DC-blocker should be designed to work between two of these transmission nulls. For more convenience we set $n = 1$. Using (4), we can plot the frequency response of $(S_{DC-blocker})_{11}$ for different values of $\left(\frac{Z_0}{Z_{CL}}\right)$ and $CR$ in the frequency range of $0 < \theta < \pi$, where $CR$ and $Z_{CL}$ are the coupling ratio and coupled-line characteristic impedance, respectively and defined by (A21) and (A22) of Appendix. Fig. 4a shows the CR=0.5 contours in the $(S_{DC-blocker})_{11}$ plane for different values of $\left(\frac{Z_0}{Z_{CL}}\right)$; and Fig. 4b shows the $\left(\frac{Z_0}{Z_{CL}}\right) = 0.5$ contours for different values of CR.

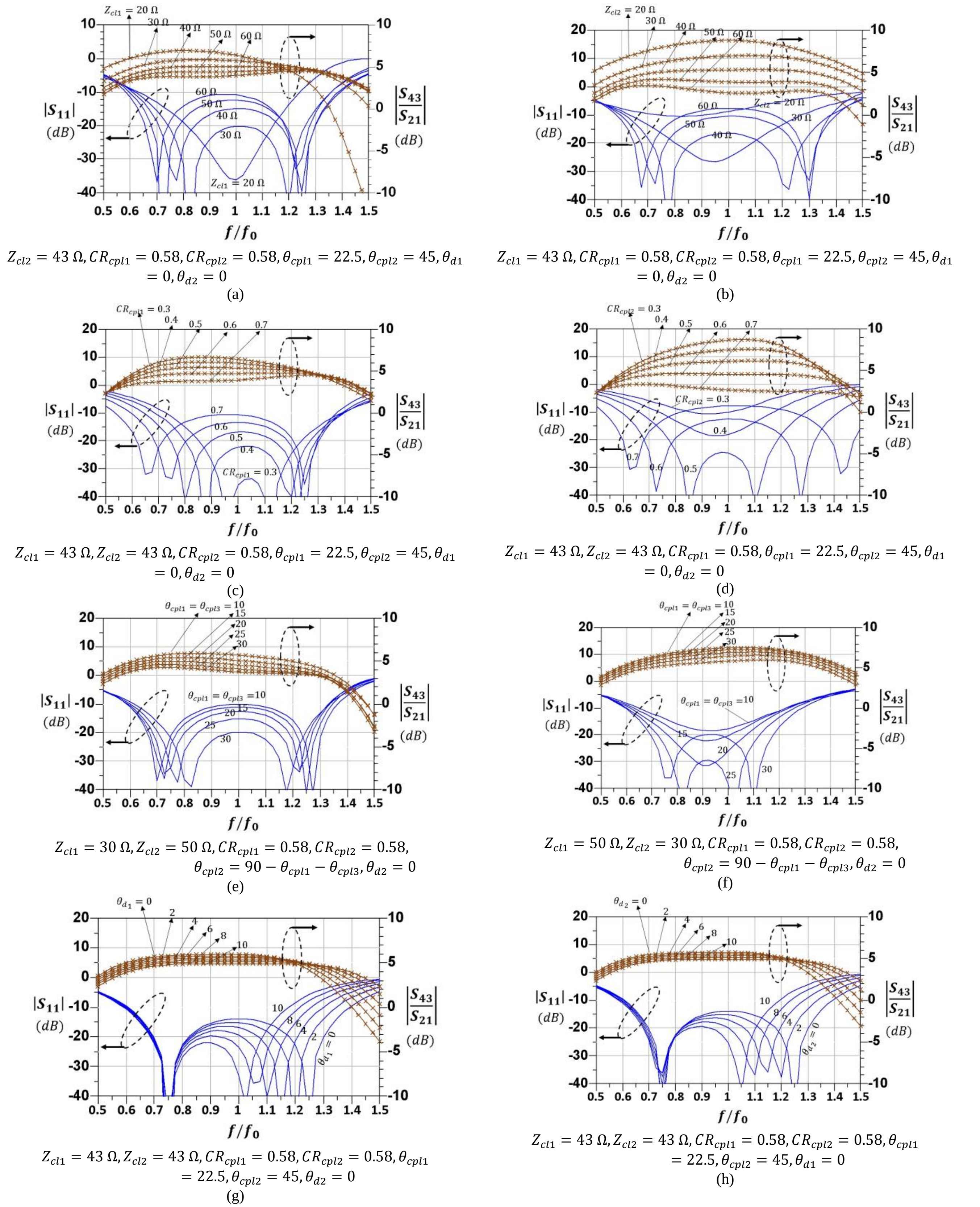

$Z_{cl2} = 43\ \Omega, CR_{cpl1} = 0.58, CR_{cpl2} = 0.58, \theta_{cpl1} = 22.5, \theta_{cpl2} = 45, \theta_{d1} = 0, \theta_{d2} = 0$

(a)

$Z_{cl1} = 43\ \Omega, CR_{cpl1} = 0.58, CR_{cpl2} = 0.58, \theta_{cpl1} = 22.5, \theta_{cpl2} = 45, \theta_{d1} = 0, \theta_{d2} = 0$

(b)

$Z_{cl1} = 43\ \Omega, Z_{cl2} = 43\ \Omega, CR_{cpl2} = 0.58, \theta_{cpl1} = 22.5, \theta_{cpl2} = 45, \theta_{d1} = 0, \theta_{d2} = 0$

(c)

$Z_{cl1} = 43\ \Omega, Z_{cl2} = 43\ \Omega, CR_{cpl1} = 0.58, \theta_{cpl1} = 22.5, \theta_{cpl2} = 45, \theta_{d1} = 0, \theta_{d2} = 0$

(d)

$Z_{cl1} = 30\ \Omega, Z_{cl2} = 50\ \Omega, CR_{cpl1} = 0.58, CR_{cpl2} = 0.58, \theta_{cpl2} = 90 - \theta_{cpl1} - \theta_{cpl3}, \theta_{d2} = 0$

(e)

$Z_{cl1} = 50\ \Omega, Z_{cl2} = 30\ \Omega, CR_{cpl1} = 0.58, CR_{cpl2} = 0.58, \theta_{cpl2} = 90 - \theta_{cpl1} - \theta_{cpl3}, \theta_{d2} = 0$

(f)

$Z_{cl1} = 43\ \Omega, Z_{cl2} = 43\ \Omega, CR_{cpl1} = 0.58, CR_{cpl2} = 0.58, \theta_{cpl1} = 22.5, \theta_{cpl2} = 45, \theta_{d2} = 0$

(g)

$Z_{cl1} = 43\ \Omega, Z_{cl2} = 43\ \Omega, CR_{cpl1} = 0.58, CR_{cpl2} = 0.58, \theta_{cpl1} = 22.5, \theta_{cpl2} = 45, \theta_{d1} = 0$

(h)

Fig. 7. Scattering parameters of the DC-blocker equivalent circuit of Fig. 6a for different design parameters values, port numbering is defined in Fig. 6b: scattering parameters as functions of: a) $Z_{cl1}$, b) $Z_{cl1}$, c) $cpl1$ coupling ratio ($CR_{cpl1}$), d) $cpl2$ coupling ratio ($CR_{cpl2}$), e) $cpl1$ electrical length ($\theta_{cpl1}$), f) $cpl2$ electrical length ($\theta_{cpl2}$), g) $delay1$ electrical length ($\theta_{d1}$), h) $delay2$ electrical length ($\theta_{d2}$).

These two figures illustrate that for a fixed CR, lower values of $\left(\frac{Z_0}{Z_{CL}}\right)$ show better frequency response in terms of the return losses. In addition, for a fixed $\left(\frac{Z_0}{Z_{CL}}\right)$, better return losses in a wider frequency band are obtained for higher values of CR. We know that higher values of CR are achieved by narrowing the dielectric gap between coupled-line conductor traces which, as explained before, introduces some practical problems. On the other hand, increasing $Z_{CL}$ (to decrease $\left(\frac{Z_0}{Z_{CL}}\right)$) is identical to using very thin conductor traces which is not a practical strategy. So, it will be desired to decrease the reference impedance value ($Z_0$). If the circuit of Fig. 3c is used as a two port network, it can be described by its impedance parameters (see Fig. 5a). A series arrangement of two networks like that of Fig. 5a is shown in Fig. 5b. According to the illustrated voltage and current signs in Fig. 5b, the excitations of two series networks have opposite signs. Because of this differential excitation, we can convert each excitation port to a series connection of two $\frac{Z_0}{2}$-ohm ports and set the voltage of their connecting node to the zero. So, the circuit is equivalent to Fig. 5c where it is shown that the scattering parameters at $Z_0$-ohm ports of the network of Fig. 5b are identical to those at $\frac{Z_0}{2}$-ohm ports in Fig. 5c. Thus, the differential mode excitation itself halves the reference impedance value. So, if we design two quadrature wavelength coupled-line structures and insert them in (+) and (-) paths of a differential-mode transmission line, we can achieve a wideband DC-blocker without any need for tightening the dielectric gap between two adjacent conductor traces or decreasing the thicknesses of the conductor traces.

## III. Designs and Simulations

### A. DC-blocker

The simplest equivalent circuit for our proposed SBCSL DC-blocker is shown in Fig. 6a. It consists of two separate similar paths of open-ended cascaded coupled-line sections ( $M_{cascaded}$s) each connects an input SBCSL conductor trace to an output SBCSL trace. It is seen in the figure that each path comprised of a cascaded connection of five separate four-port networks, where the first, third and last networks (namely *cpl1*, *cpl2* and *cpl3*) are coupled-line sections and *delay1* and *delay2* are isolated delay lines. Since the figure implies that *cpl3* and *delay2* are rotated copies of *cpl1* and *delay1*, respectively, their corresponding scattering transfer matrices are equal to:

$$M_{Total}^{3} = E_4 (M_{Total}^{1})^{-1} E_4 \quad (9)$$

$$M_{delay}^{2} = E_4 \left(M_{delay}^{1}\right)^{-1} E_4 \quad (10)$$

where

$$E_4 = \begin{pmatrix} 0 & 0 & 0 & 1 \\ 0 & 0 & 1 & 0 \\ 0 & 1 & 0 & 0 \\ 1 & 0 & 0 & 0 \end{pmatrix} \quad (11)$$

So, $4 \times 4$ scattering transfer matrix of the cascaded coupled-line sections is derived as

$$\begin{aligned} M_{cascaded} = M_{Total}^{1} \times M_{delay}^{1} \times M_{Total}^{2} \\ \times E_4 \left(M_{delay}^{1}\right)^{-1} E_4 \\ \times E_4 (M_{Total}^{1})^{-1} E_4 \end{aligned} \quad (12)$$

In (12) $M_{Total}^{1}$ and $M_{Total}^{2}$ are calculated by (2) and $M_{delay}^{1}$ is simply equivalent to

$$M_{delay}^{1} = \begin{bmatrix} e^{j\theta_{d_1}} & 0 & 0 & 0 \\ 0 & e^{j\theta_{d_2}} & 0 & 0 \\ 0 & 0 & e^{-j\theta_{d_1}} & 0 \\ 0 & 0 & 0 & e^{-j\theta_{d_2}} \end{bmatrix} \quad (13)$$

where $\theta_{d_1}$ and $\theta_{d_2}$ are the electrical lengths of the depicted delay lines.

The differential mode and common mode equivalent circuits of the DC-blocker are shown in Fig. 6b. It is seen in the figure that in the differential mode, the DC-blocker is approximately equivalent to the series connection of two $M_{cascaded}^{2D}$ networks and a phase-inverter block. $M_{cascaded}^{2D}$ can be derived from $M_{cascaded}$ using the same procedure that we calculated $M_{DC-blocker}$ from $M_{Total}$ (look (3) and (4)). Also Fig. 6b illustrates that in the common mode, the DC-blocker is approximately equivalent to the parallel connection of two $M_{cascaded}^{2D}$ networks.

According to (A20) - (A23) cpl1, cpl2 and cpl3 sections are characterized by their electrical lengths ($\theta_{cpl}$s), coupling ratios (CRs), matched impedance of the coupled-line sections ($Z_{cl}$s) and the reference impedance ($Z_0$). As we later show, the depicted delay lines are added to the structure only for practical issues and they are assumed matched to $Z_0$ in our primary design.

The scattering parameters of the circuit of Fig. 6a as a function of different circuit parameters are plotted in Figs. 7a-7h. The scattering parameters are calculated at port1, port2 of the differential mode circuit and port3 and port4 of the common mode circuit as depicted in Fig. 6b. In each figure, two groups of curves are plotted which are namely $|S_{11}|$ (in differential mode) and $\frac{|S_{21}|}{|S_{43}|}$ which shows the selectivity of the DC-blocker between the differential mode and common mode signals (or approximately the common mode rejection ratio CMRR). All curves have been plotted as a function of normalized frequency ($\frac{f}{f_0}$). $f_0$ is the frequency that the electrical length of the coupled-line sections is equal to $\frac{\pi}{2}$. It is seen from the figures that one can achieve about 66% of relative bandwidth for a differential mode input return loss of better than 15 dB and $\frac{|S_{21}|}{|S_{43}|}$ better than 5 dB. Figs. 7a, 7b imply that there is a tradeoff between $|S_{11}|$ bandwidth and $\frac{|S_{21}|}{|S_{43}|}$ level for different values of $Z_{cl1}$ and $Z_{cl2}$. This tradeoff is seen in Figs. 7c, 7d for different $CR_{cpl1}$ and $CR_{cpl2}$ values, too. Figs. 7e, 7f show that the electrical lengths of the coupled-line sections (when their total electrical length is equal to $\frac{\pi}{2}$ ) only affect the input return loss of the DC-blocker. Finally, it is seen in Figs. 7g, 7h that both the $|S_{11}|$ bandwidth and $\frac{|S_{21}|}{|S_{43}|}$ bandwidth are shortened when the electrical lengths of the delay lines ($\theta_{d1}$ and $\theta_{d2}$) increase.

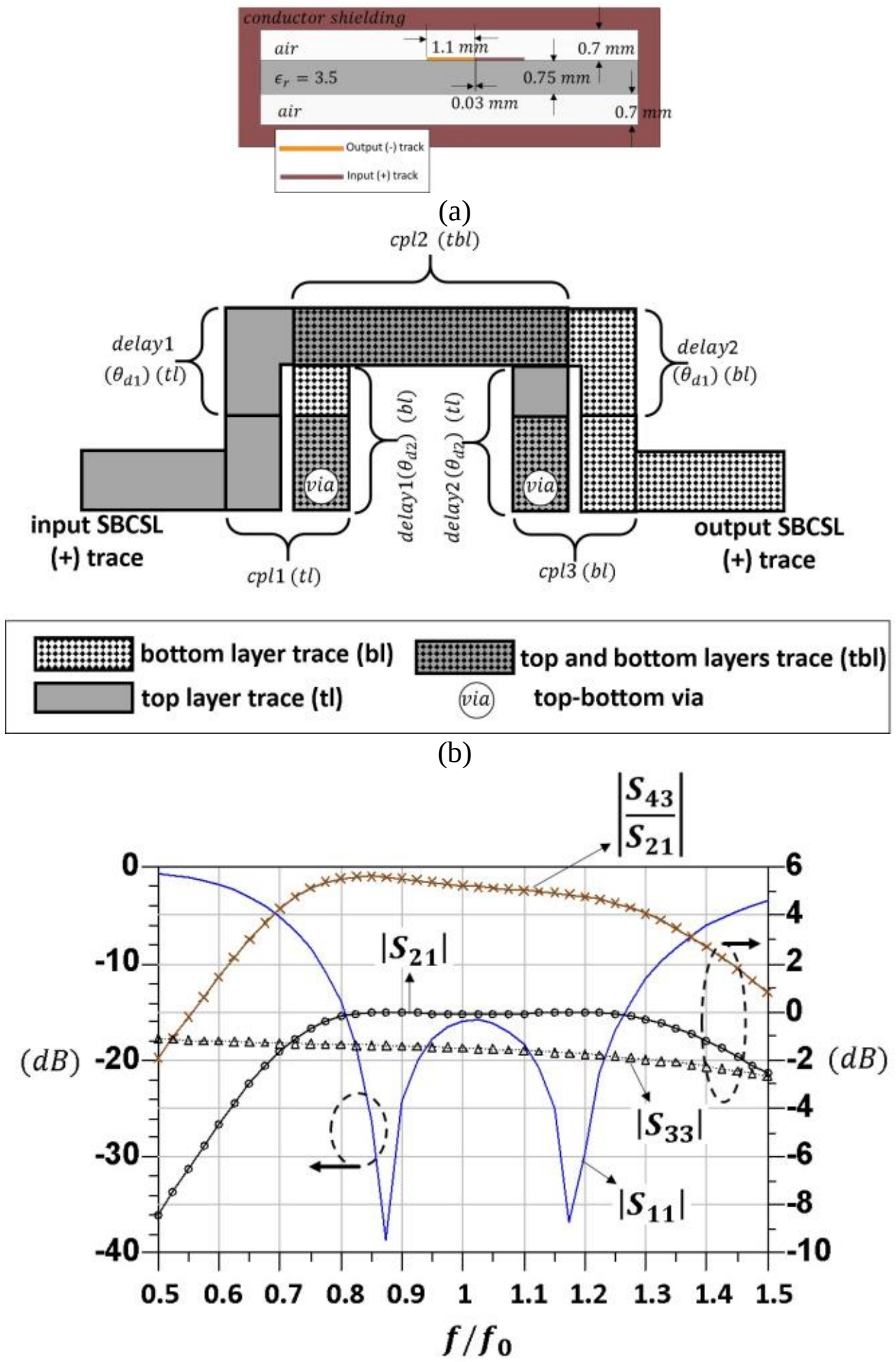

Fig. 8. Single-layer PCB realization of the DC-blocker of Fig. 6a: a) $cpl1$ and $cpl2$ unoptimized realization of Table. I, b) our proposed optimized design, c) scattering parameters of the optimized design for values listed in Table. II.

TABLE I
UNOPTIMIZED DESIGN VALUES FOR THE COUPLED-LINE SECTION OF FIG. 8A

| $Z_{CL1}$ | $Z_{CL2}$ | $CR_{cpl1}$ | $CR_{cpl2}$ | $\theta_{cpl1} = \theta_{cpl3}$ | $\theta_{cpl2}$ | $\theta_{d1} = \theta_{d2}$ | $l$ (length) |
|---|---|---|---|---|---|---|---|
| 43.3 Ω | 43.3 Ω | 0.58 | 0.58 | $22.5^o$ | $45^o$ | $o$ | $\frac{\lambda_g}{4}$ @ $f_0$ |

TABLE II
DESIRED VALUES FOR THE DC-BLOCKER OF FIG. 8B.

| $Z_{CL1}$ | $Z_{CL2}$ | $CR_{cpl1}$ | $CR_{cpl2}$ | $\theta_{cpl1} = \theta_{cpl3}$ | $\theta_{cpl2}$ | $\theta_{d1}$ | $\theta_{d2}$ |
|---|---|---|---|---|---|---|---|
| 100 Ω | 65 Ω | 0.3 | 0.75 | $14.4^o$ | $22.4^o$ | $12^o$ | $16^o$ |

Now, we consider the case of an unoptimized DC-blocker with design parameters listed in Table. I. According to Fig. 7a such a DC-blocker shows a wideband behavior around the center frequency of $f_0$ (about 66% of relative bandwidth with return losses of better than 15 dB). The cross section of a typical realization of *cpl1* and *cpl2* with specifications replaced by values in Table I for a single-layer PCB is shown in Fig. 8a. In contrast to the IC realization which is practically feasible, the PCB realization has some limitations due to the quasi-TEM wave propagation which does not have any analytical solution. The even and odd modes of a single-layer PCB coupled-line structure have different effective permittivities and so experience different phase velocities. Since, there is no analytical expression for the even and odd modes impedances with respect to the dimensions of the coupled-line structure, these impedances should be empirically calculated like those of [22]. As shown in Fig. 8a, the metallic traces should be very close to each other (0.03 mm). This tight spacing of metallic traces is not practically realizable. Even if one can realize this narrow gap, there is a high risk of dielectric breakdown in the case of unequal DC voltages of the traces. Also, the unnecessary 1.1- mm traces widths and low 0.7-mm distances from top and bottom shielding are the other disadvantages of this realization.

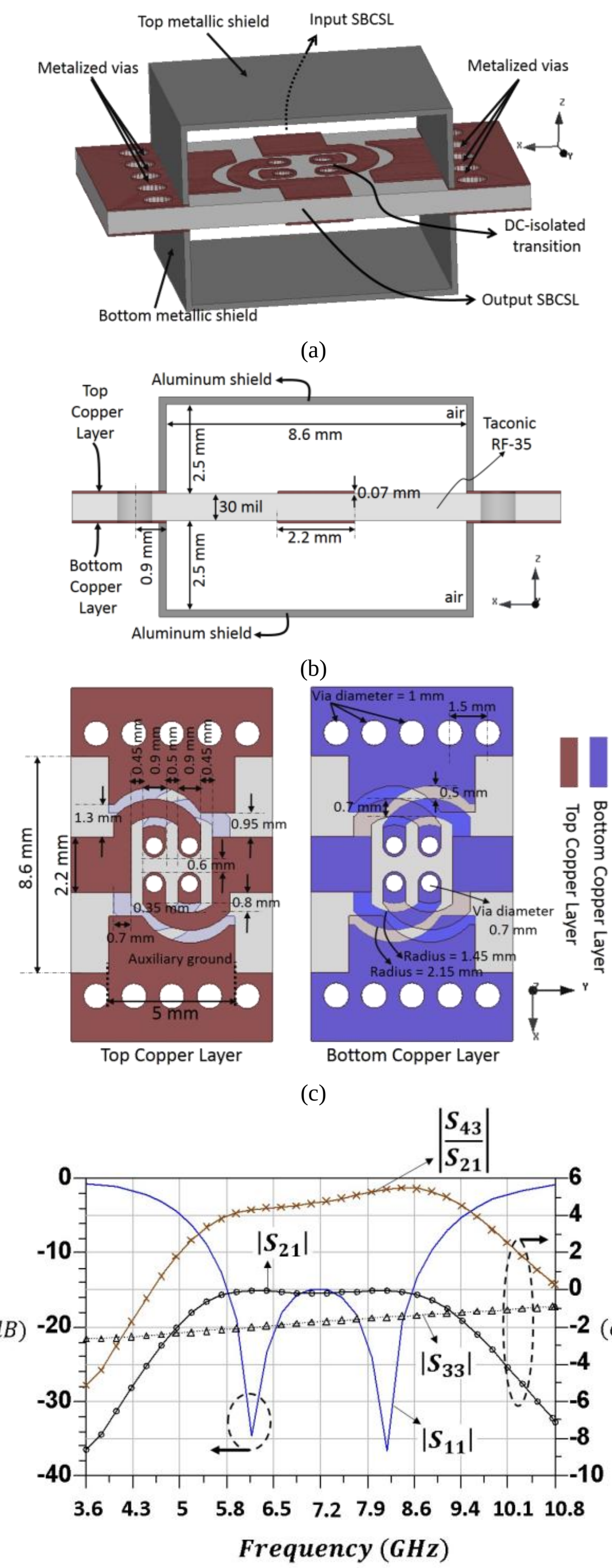

Fig. 9. Optimized DC-blocker: a) single-layer PCB realization of the component for SBCSL structure, b) input and output SBCSL structures cross sections, c) detailed specifications of the transition, d) scattering parameters based on the ports definition of Fig. 6b.

To avoid the mentioned restrictions, we have designed our DC-blocker in the form illustrated in Fig. 8b. This figure shows a half of the DC-blocker which is a path between the input SBCSL (+) trace and the output SBCSL (-) trace following Fig. 6a definitions. The other half has the same but rotated structure. The figure shows that $cpl1$ and $cpl3$ are top-layer and bottom-layer edge-coupled coupled-line structures, respectively. But, $cpl2$ is a top-bottom broadside-coupled coupled-line structure which allows the coupling ratio ($CR_{cpl2}$) to be increased to the desired value. In addition, $delay1$ consists of two separate delay sections: a top-layer trace ($\theta_{d1}$), and a connection of a metallized via and a bottom-layer trace ($\theta_{d2}$). It is obvious from the figure that this configuration already features the phase inversion property. The design values of the optimized DC-blocker with the configuration of Fig. 8b is listed in Table. II. These values are obtained for the best return loss and common mode rejection performance. It is worth mentioning that a big restriction in our design is the values of $\theta_{d1}$ and $\theta_{d2}$ which for the sake of feasibility and fabrication limitations cannot be decreased very much. We remember from Fig. 7g, 7h that these two parameters seriously affect the design bandwidth. However, Table. II shows that the total electrical length of the proposed DC-blocker is approximately 80 deg which is 10 deg smaller than that of a simple quadrature-length structure. Fig. 8c shows the scattering parameters of the optimized design. It is shown that better than 10 dB of return loss is obtained in a 50% of relative bandwidth. Also the common mode rejection level is better than 5 dB in the same frequency band.

The optimized DC-blocker configuration is simulated in Ansoft HFSS (Fig. 9a). It is connected to the input and output SBCSL structures at both sides. The final structure is finely tuned in Ansoft HFSS for the best possible performance. The shown metalized vias complete the outer metallic shield inside the Taconic-30mil-RF35 laminate. The cross section of the input/output SBCSLs (as illustrated in Fig. 9b) are composed of two broadside-coupled 2.2-mm×70-$\mu m$ copper traces spaced 2.5 mm from the top and bottom aluminum shields. In this structure $Z_0 = 50\Omega$ for differential mode propagation. The thorough depiction of the structure in Fig. 9c (compare to the model of Fig. 8b) shows that there is a 0.35-mm airgap between the coupled traces. This is a noticeable modification to the mentioned 0.03-mm. The traces have 0.7mm widths and spaced 0.5 $mm$ from the auxiliary grounds. Fig. 9c shows that the top and bottom layers are symmetric but mirrored with respect to each other. The shown 1-mm ground vias and 0.7-mm traces vias are both easy to fabricate. There is not any airgap smaller than 0.35 $mm$ in the shown structure. The connecting SBCSL conductor traces of 2.2-mm width are also shown in the figure. The simulated S-parameters of the transition at the input and output 50-ohm SBCSL ports are shown in Fig. 9d. Assuming $f_0 = 7.2\ GHz$ the parameters are plotted in the same scale as that of Fig. 8c. The figure implies that the simulated structure shows frequency responses very close to that of the ideal structure. A return loss of better than 15 dB for a bandwidth of 2.8 GHz (39 %) and 10 dB for a bandwidth of 3.2 GHz (45 %) have been achieved. Also, the common mode rejection level is better than 4 dB in the same 50 % of relative bandwidth. The overall length of the transition is 5 mm which is equal to $\frac{\lambda_g}{5}$ at 7.2 GHz. This length includes the required interconnects and additional airgaps between the open-end coupled-line sections and the input / output SBCSLs.

TABLE III

VARIATION OF THE NORMALIZED CENTER FREQUENCY ($\frac{f_c}{f_c^0}$) AS A FUNCTION OF PERTURBATIONS IN THE WIDTH AND LENGTH OF THE DC-BLOCKER

| | $\alpha_w = 0.9$ | $\alpha_w = 0.95$ | $\alpha_w = 1.05$ | $\alpha_w = 1.1$ |
|---|---|---|---|---|
| $\alpha_L = 0.9$ | 1.07 | 1.06 | 1.05 | 1.04 |
| $\alpha_L = 0.95$ | 1.04 | 1.02 | 1 | 1 |
| $\alpha_L = 1.05$ | 1 | 0.98 | 0.97 | 0.96 |
| $\alpha_L = 1.1$ | 0.98 | 0.96 | 0.94 | 0.93 |

TABLE IV

VARIATION OF THE NORMALIZED 10-DB RETURN LOSS BANDWIDTH ($\frac{BW}{BW_0}$) AS A FUNCTION OF PERTURBATIONS IN THE WIDTH AND LENGTH OF THE DC-BLOCKER

| | $\alpha_w = 0.9$ | $\alpha_w = 0.95$ | $\alpha_w = 1.05$ | $\alpha_w = 1.1$ |
|---|---|---|---|---|
| $\alpha_L = 0.9$ | 1 | 1.04 | 1.09 | 1.125 |
| $\alpha_L = 0.95$ | 0.94 | 1 | 1.05 | 1.09 |
| $\alpha_L = 1.05$ | 0.82 | 0.92 | 1 | 1.05 |
| $\alpha_L = 1.1$ | 0.74 | 0.83 | 0.92 | 1 |

It is worth testing the sensitivity of the designed phase-inverted DC-blocker with respect to dimension changes. Since the DC-blocker has too many dimension parameters, we define a width scale factor ($\alpha_w$) and a length scale factor ($\alpha_L$) and calculate the effect of these two factors on the value of the DC-blocker center frequency ($f_c$) and 10-dB return loss bandwidth ($BW$). The original DC-blocker has the center frequency of $f_c^0 = 7.2\ GHz$ and 10-dB bandwidth of $BW^0$ (5.6-8.8 GHz), as shown in Fig. 9d. Also, the width and length of the original design are $W_0$ and $L_0$, respectively. We have simulated several DC-blockers with perturbed widths ($\alpha_W W_0$) and lengths ($\alpha_L L_0$) and listed the values of their corresponding normalized center frequency ($\frac{f_c}{f_c^0}$) and normalized 10-dB return loss bandwidth ($\frac{BW}{BW_0}$) in Table. III and Table. IV, respectively. It can be infered from Table. III that increasing either the DC-blocker width or its length will decrease $f_c$ but with different scales. However, as shown in Table. IV increasing DC-blocker width and length have inverse effects on the 10-dB return loss bandwidth ($BW$). Wider bandwidths can be achieved by either increasing DC-blocker width or decreasing its length. Also, for a same values of width and length scales $BW$ remains unchanged. Moreover, it is obvious from Table. III and Table. IV that $BW$ is more sensitive to dimension changes in comparison to $f_c$.

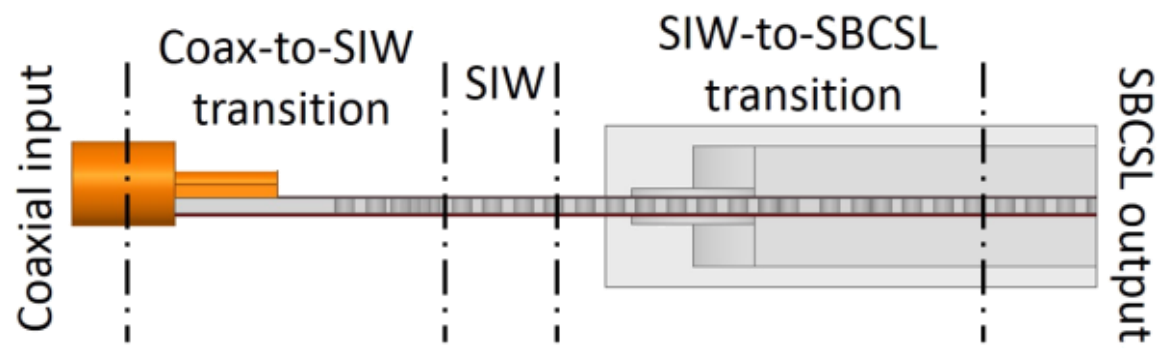

(a)

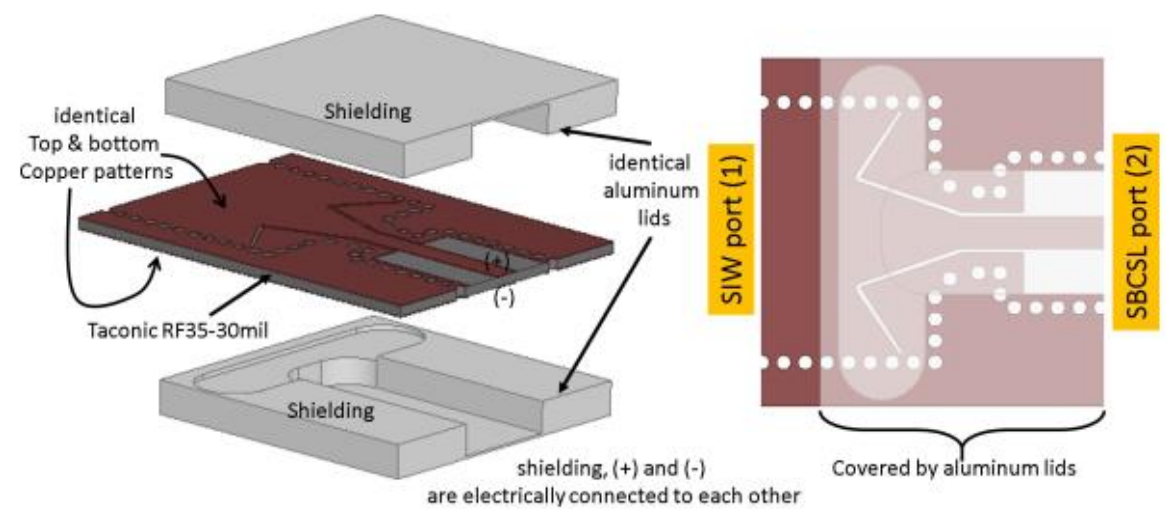

(b)

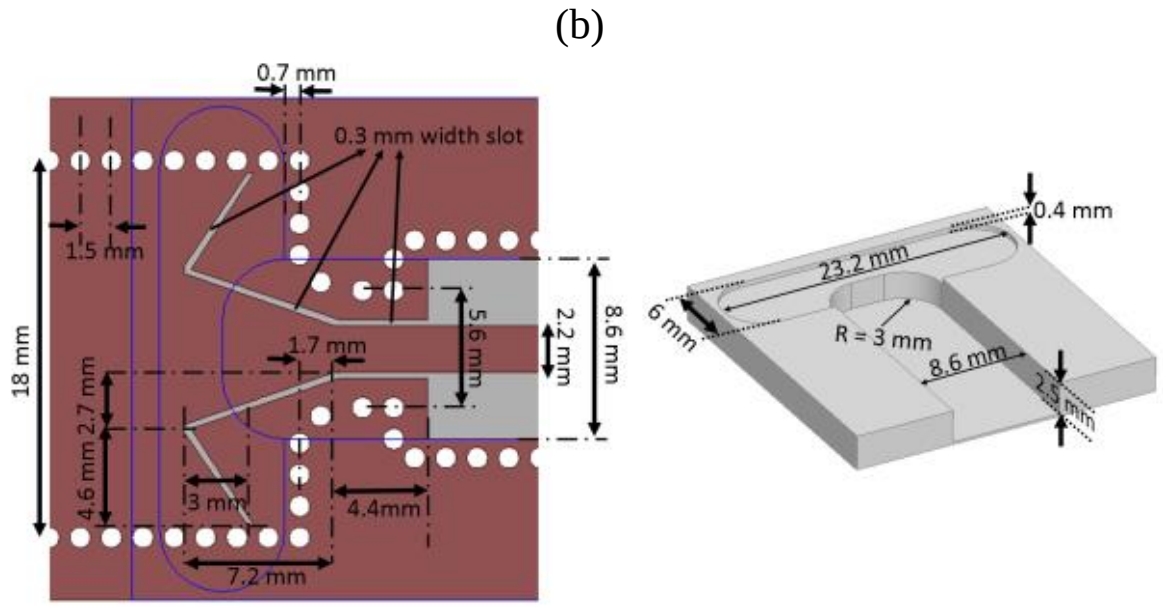

(c)

Fig. 10. Proposed novel SIW-mediated balun: a) overall configuration, b) explosive view of the constitutive parts, c) dimensions,

### *B. SIW-mediated Balun*

For the measurement of the balanced DC-blocker S-parameters, we have designed a wideband SIW-mediated balun. This component is expected to properly transform the unbalanced coaxial TEM single-ended mode into the balanced SBCSL quasi-TEM differential mode. The proposed balun configuration is illustrated in Fig. 10a. It simply consists of a coaxial-to-SIW and SIW-to-SBCSL transitions. The main role of the intermediate SIW section is to electrically connect all SBCSL conductors together, namely (+), (-) and the shielding. The SBCSL structure is composed of two copper traces printed on the top and bottom layers of a Taconic-RF35-30mil laminate (with $\epsilon_r = 3.5$ and $\tan\delta$=0.0018), while the whole PCB is surrounded by two aluminum lids (Fig. 10b). The transition contains of two slots on the top and bottom layers which gradually transforms the SIW guiding mode into the differential mode of the SBCSL. Since the slots are entirely covered by the aluminum plates air cavities, there will be no chance for any radiation loss mechanisms. Also, due to the very small height of the cavities, no resonant modes are excited and probable resonances and total power reflections are all entirely eliminated. It is worth to mention that, the top and the bottom copper layers of the PCB have identical patterns. The dimensions of the SBCSL transmission line are chosen similar to those of the DC-blocker of Fig. 8b. Also, the dimensions of the SIW ( like $SIW\ width \approx 18\ mm$) are chosen in such a way that in the entire frequency range of 5.6-8.8 GHz (which is our desired transition band) it operates in its single-mode condition. The detailed dimensions of the transition (PCB and aluminum lids) are found in Fig. 10c. In this figure it is seen that the transition slots are placed inside $0.4\ mm \times 6\ mm \times 23.2\ mm$ top and bottom cavities. The total length of the transition is about 13 mm. The S-parameters of the transition between the SIW and the SBCSL ports (highlighted in Fig. 10b) are plotted in Fig. 11a which is designed and simulated in Ansoft HFSS. As can be seen, from slightly above the SIW cutoff frequency (4.7 GHz) the return loss exceeds 20 dB and remains above this value in the entire SIW single-mode band. The $|S_{21}|$ and $|S_{11}|$ figures show no signs of radiation losses or total power reflections or resonant-like behavior. A completely flat response for $|S_{21}|$ is achieved above 5 GHz which is definitely novel in the literature. The E-field intensity distribution is plotted in the same figure which shows proper transition of the electromagnetic wave inside the structure. Also, the E-field vector distribution at the symmetry plane shows that the transition has properly excited the propagating differential mode of the SBCSL structure. The simulation results of the proposed balun when is used as divider are depicted in Fig. 11b. More than 40 dB of common mode rejection ratio (CMRR) is achieved above 5 GHz. Also the phase and amplitude imbalance are both less than 0.5% above 5 GHz. The simulations show that the balun has an excellent behavior as a 180-degree divider, too.

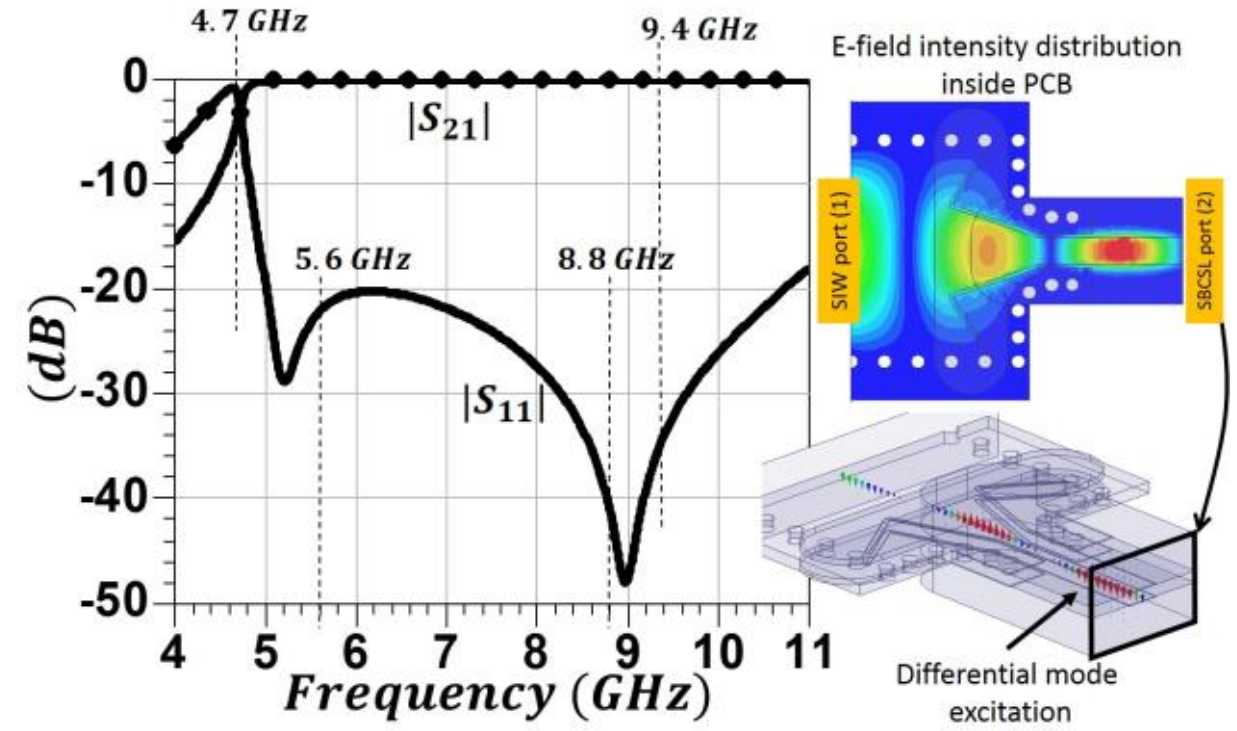

(a)

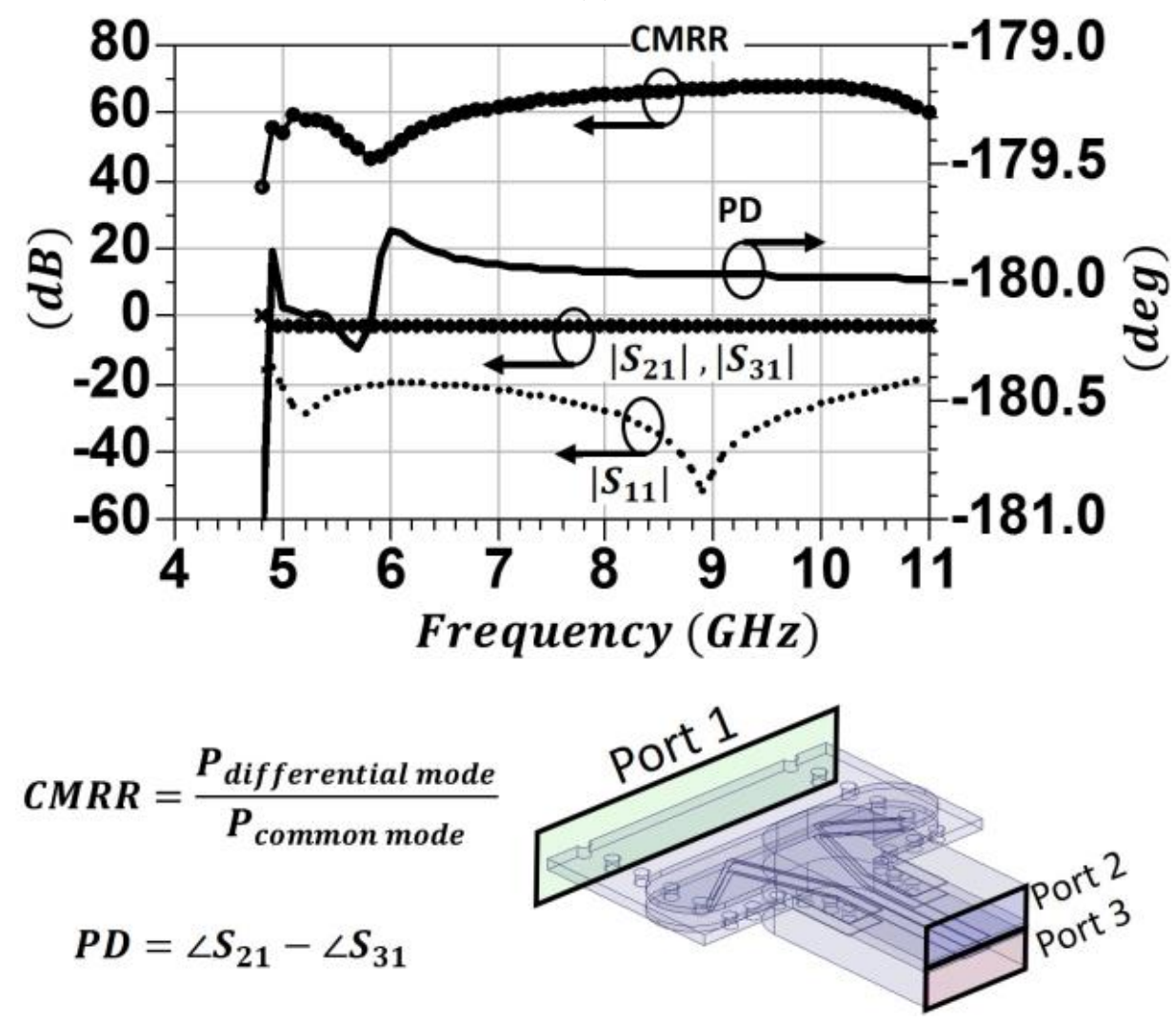

(b)

Fig. 11. HFSS simulation of the proposed SIW-mediated balun: a) its scattering parameters, b) balun common-mode rejection ratio and other figures of merit.

TABLE V
COMPARISON BETWEEN THE PROPOSED BALUN PERFORMANCE WITH SOME OTHER FAMILIAR SOLUTIONS PRESENTED IN FIG. 12.

| | Input/output matching | Radiation loss |
|---|---|---|
| solution1 | good | weak |
| solution2 | weak (resonant behavior) | good |
| solution3 | weak (narrow band) | good |
| our work | good | good |

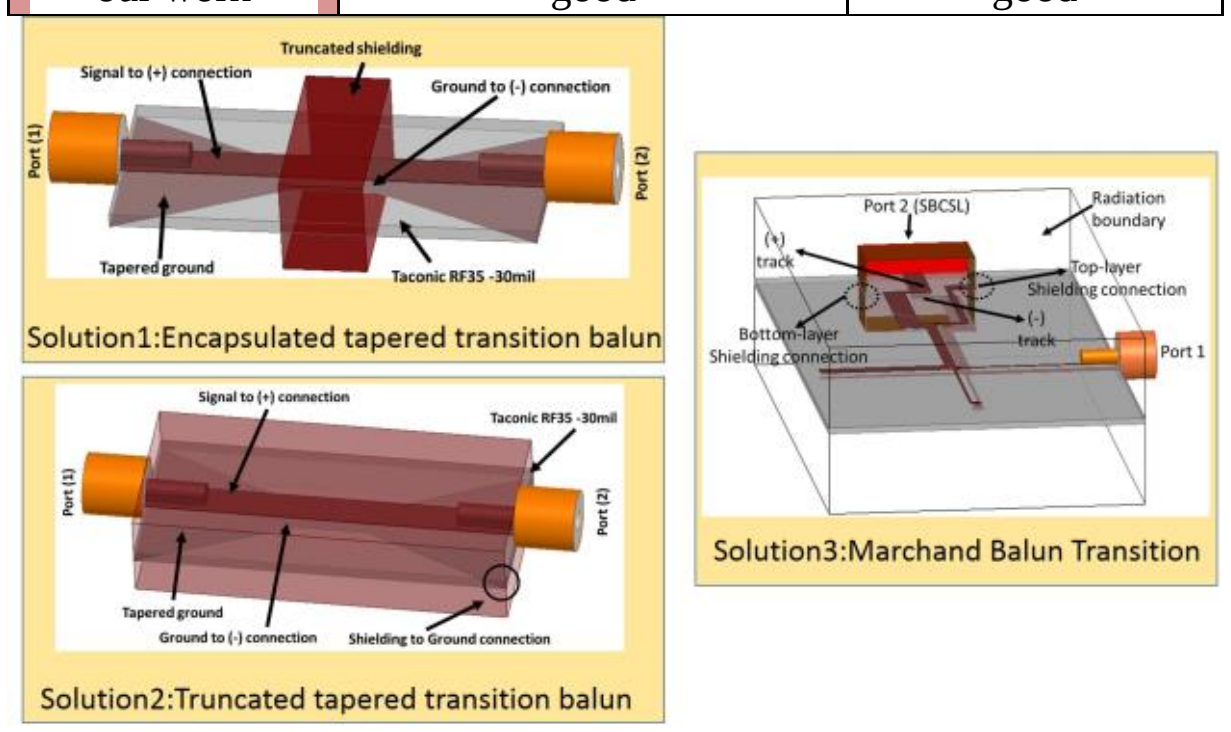

Fig. 12. Comparison of our proposed balun performance with other available solutions.

A brief comparison is able to clarify the advantage of the presented structure over other coupled-line balun solutions. To compare, we consider three different solutions (Fig. 12): the shielded parallel plate balun with tapered transition stages [23] (solution1), the parallel plate balun with tapered transition stages and truncated shielding (solution2), and the coaxial-to-SBCSL Marchand balun transition (solution3) [24]. These solutions were compared with our design with respect to the level of the undesired radiation and the input/output matchings levels. As listed in Table. V, the first solution suffers from high levels of radiation losses due to the closeness of the shielding to the tapered sections. In the second solution, single-frequency total power reflections at the input and output ports does not allow the balun to have a wideband behavior. Also, solution3 shows a narrow-band behavior when the familiar Marchand balun is used as a one input one output balun transition. As we discussed earlier, in our proposed balun the level of the undesired radiation is very low and the input and output return losses show non-resonant wideband behaviors.

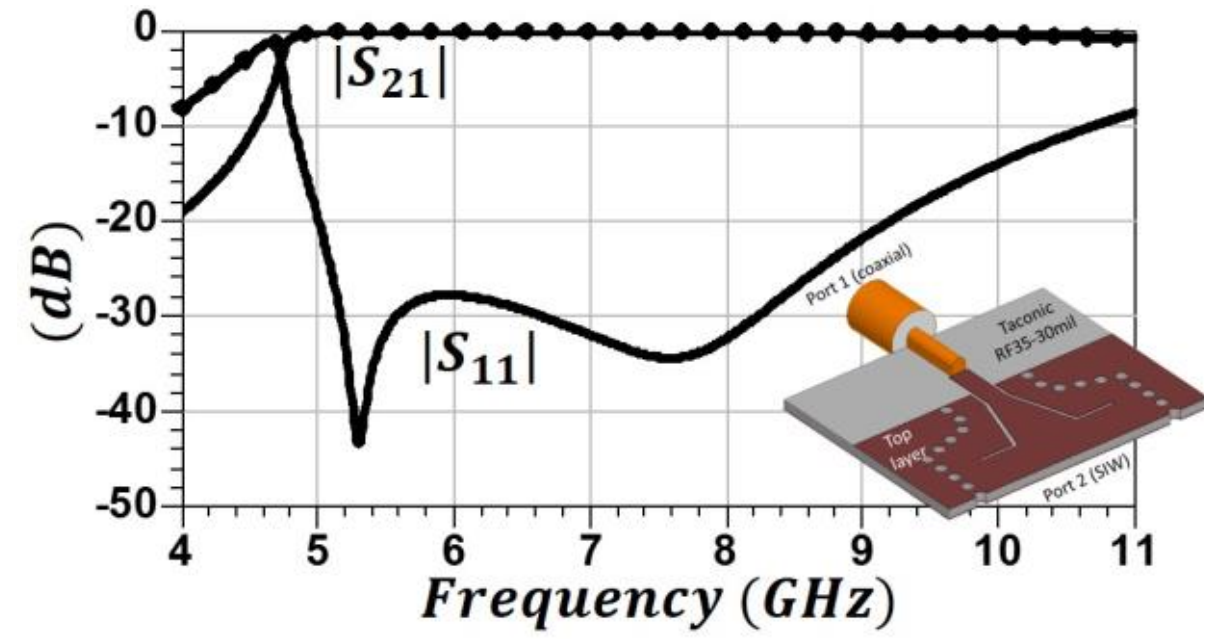

Fig. 13. Designed coaxial-to-SIW transition and its S- parameters.

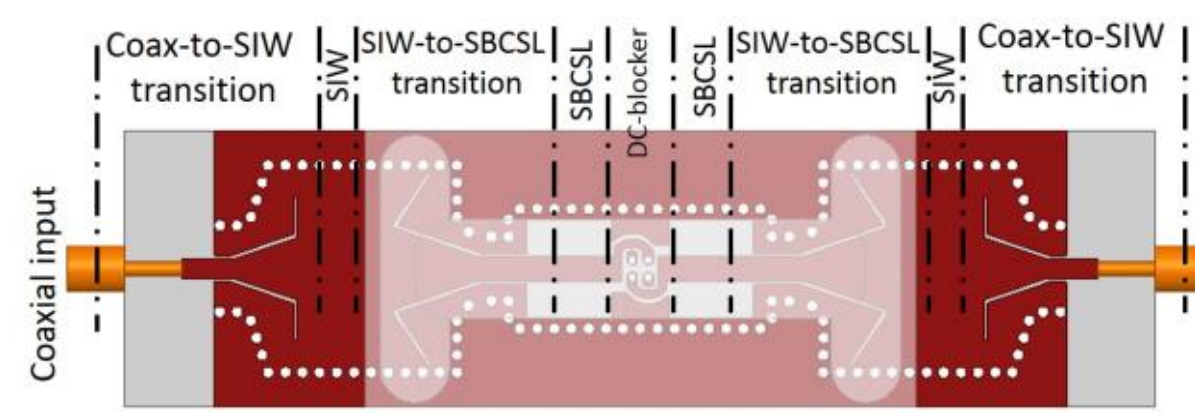

Fig. 14. Schematic of the DC-blocker prepared for the measurement by means of balun transitions and coax-to-SIW connectors.

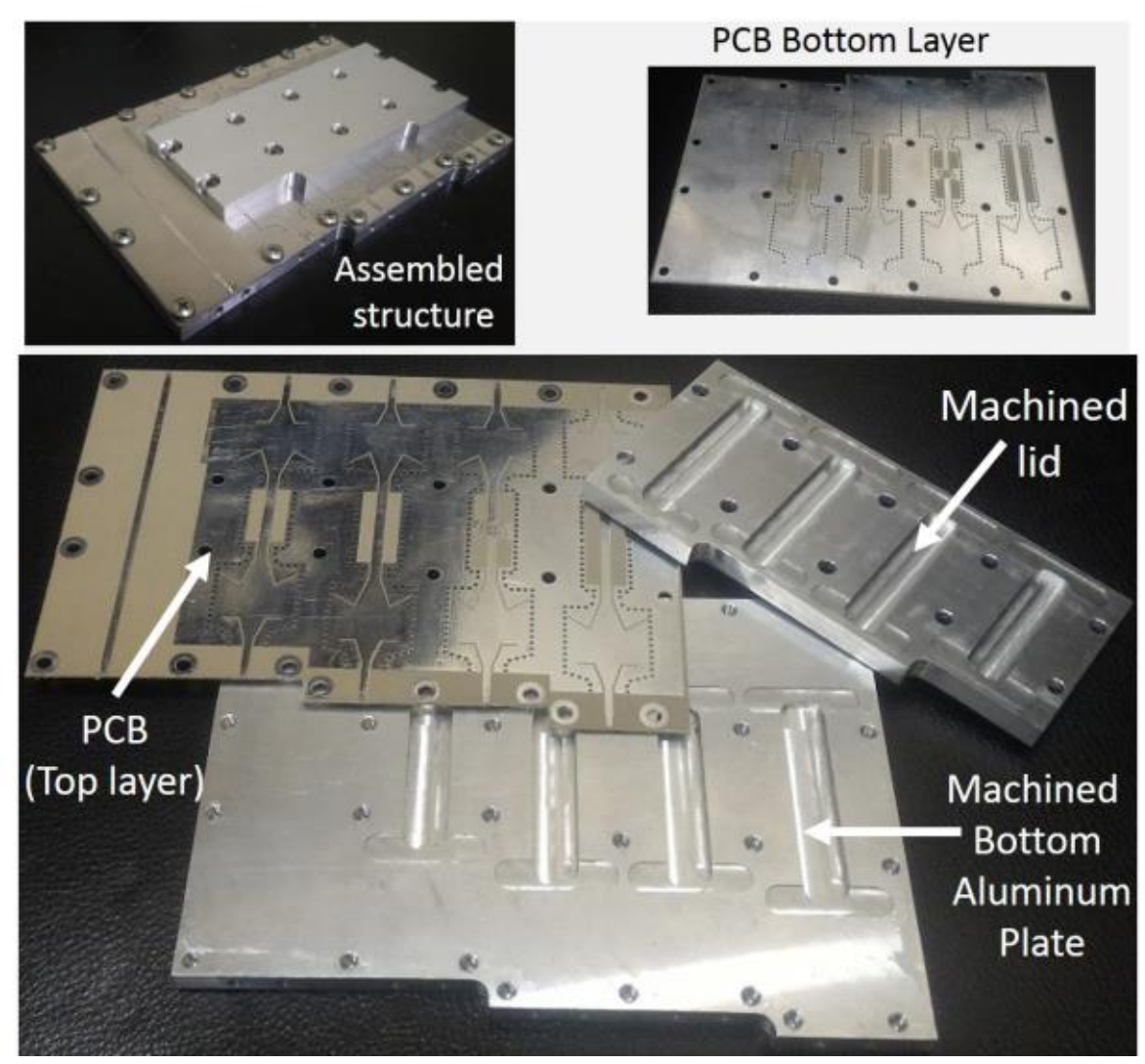

Fig. 15. Photo of the fabricated sample, assembled and de-assembled views.

## IV. Fabrication and Measurement

The scattering parameters of the coax-to-SIW transition of Fig. 10a is shown in Fig. 13. The design and optimization of this transition is based on the CPW-to-SIW transition proposed in [25]. The figure shows that the return loss of better than 28 dB is obtained for this transition in the frequency range of 5.2 – 8.4 GHz. More to say, $|S_{21}|$ shows a smooth and non-radiative behavior in the frequency region of interest. Next, for the sake of measurement, the designed coaxial-to-SIW transition, SIW-to-SBCSL balun transition and the SBCSL DC-blocker are all attached together in a series and back-to-back fashion as shown in Fig. 14.

The fabricated circuit is shown in Fig. 15. It consists of a Taconic RF-35 30mil laminate which is patterned on both sides. The circuit is assembled on a machined aluminum plate (Bottom plate). Another machined aluminum plate (lid) is placed on top of the PCB and is firmly attached to the PCB and the bottom plate. Also, the bottom side of the PCB and the total assembled structure are shown in the figure. The top lid and the bottom plate construct the required closed cavities for the SBCSL structure, DC-blocker and the balun transitions. For each S-parameters measurement process, the high frequency SMA connectors will be attached to the assembly from sidewalls.

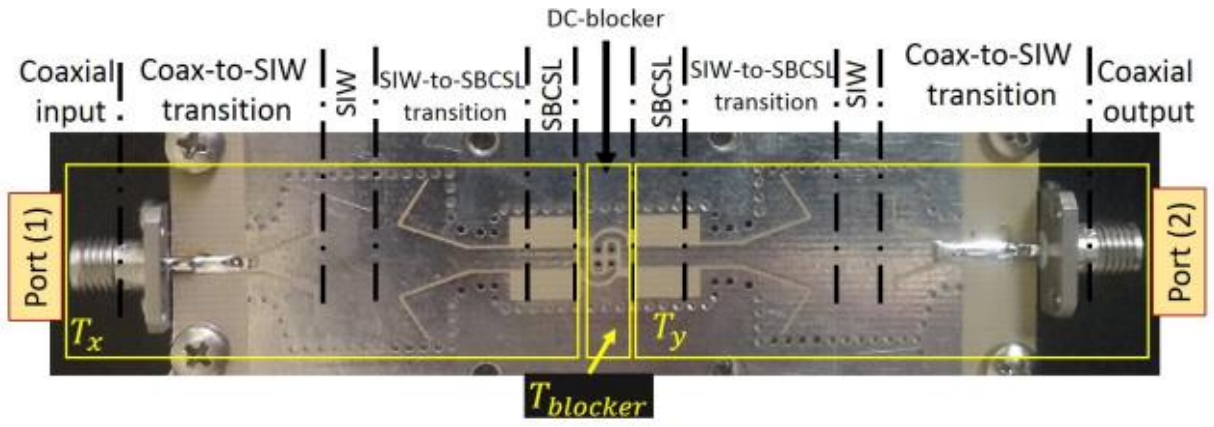

(a)

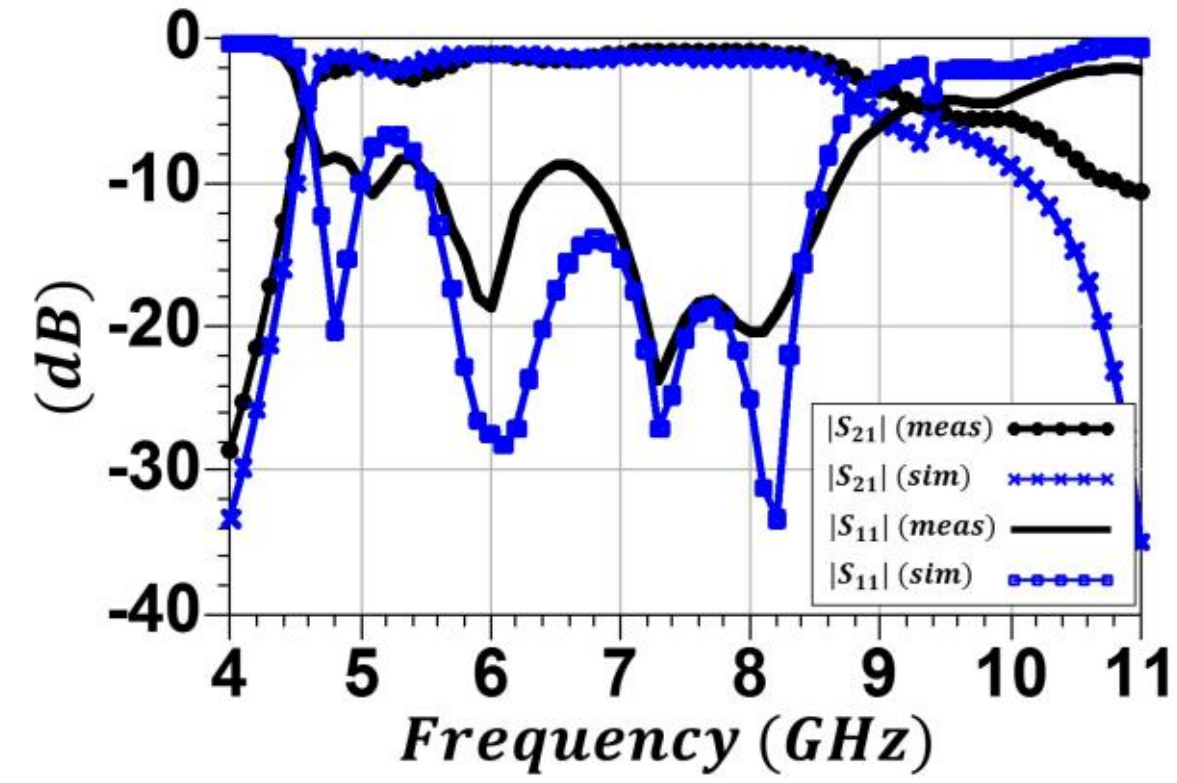

(b)

Fig. 16. The fabricated DC-blocker with its connecting baluns. A) photo of the structure, b) its S-parameters.

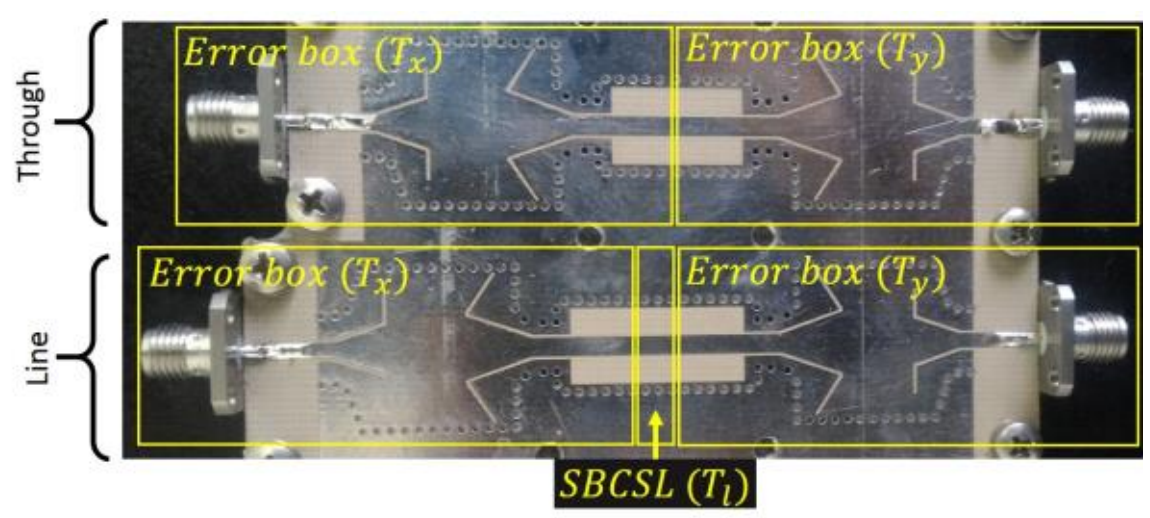

(a)

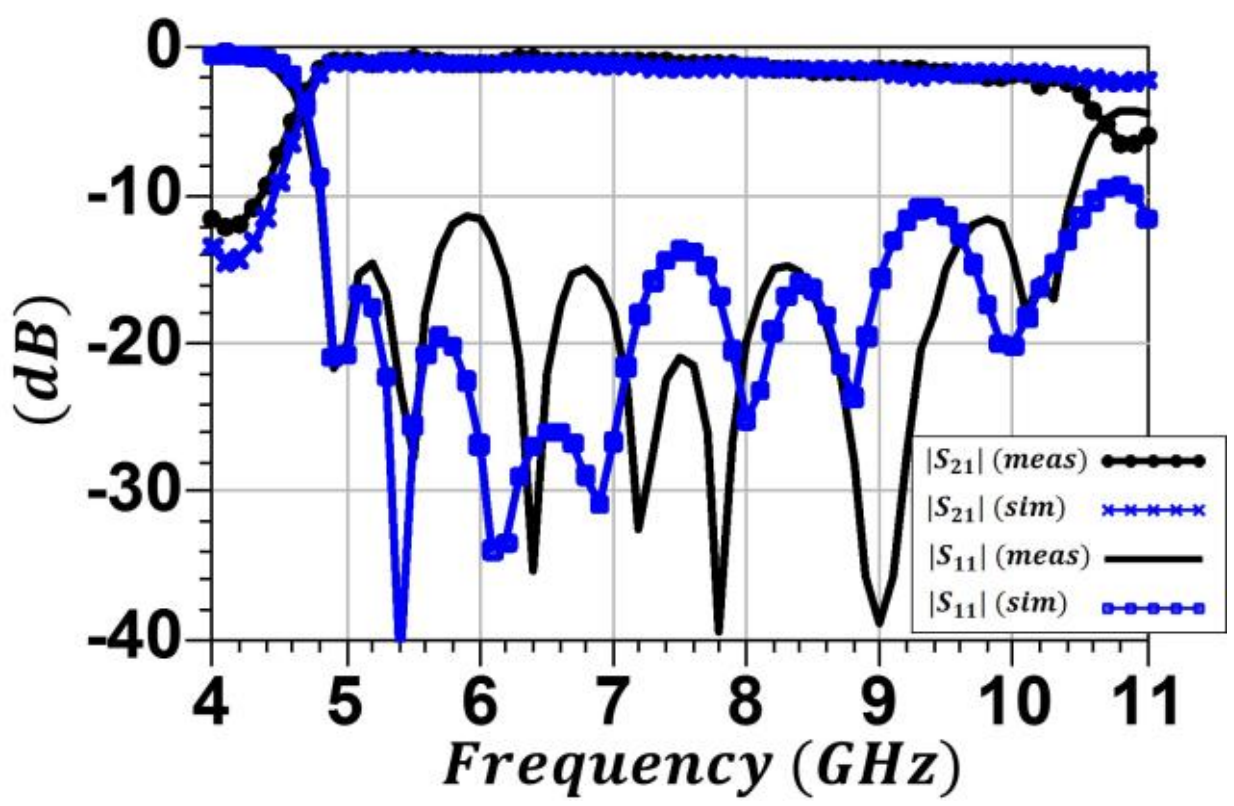

(b)

Fig. 17. The calibration kit: a) photo of the calibration kit, b) S-parameters of Line circuit.

We have designed a simple 50 Ω microstrip line on the PCB (it can be seen in Fig. 15) to model the dielectric and conductor losses in our simulations. We have used a piecewise linear model for the utilized dielectric to precisely compensate for higher values of losses seen in our measurements. With this modified dielectric model, 0.3 dB of difference is seen between the simulated and measured $|S_{21}|$ of the microstrip line in the entire 5.6-8.8 GHz frequency band.

Fig. 16a shows the fabricated sample of the structure of Fig. 14. The top aluminum lid is removed in Fig. 16a for a better illustration. The S-parameters are measured at the depicted coaxial ports. The measured and simulated results are shown in Fig. 16b. It is seen that the measurement and simulation results have similar behavior in the entire frequency band. The cutoff frequency and the nulls of $|S_{11}|$ are excellently predicted in the simulations. Even the upper band frequency responses of both simulation and measurement are approximately similar. Except for a very narrow frequency band around 6.5 GHz, $|S_{11}|$ is better than 10 dB in 5.6-8.4 GHz for the simulations and measurements. This means the relative bandwidth is 40% for the center frequency of 7 GHz. Also, the measured insertion loss is less than 1.6 dB in 5.7-8.4 GHz frequency band which its difference with that of simulation is less than 0.4 dB in the entire 5.9-8.4 GHz band.

Since both the DC-blocker and the SIW-based balun transition contribute the results shown in Fig. 16b, we have fabricated an auxiliary calibration kit to separately extract S-parameters of each component. This kit (as shown in Fig. 17a) consists of Line and Through circuits which are almost identical except for the length of the SBCSL connecting section. Before advancing with the calibration, it is worth looking at the S-parameters of the kit itself. Line circuit S-parameters are depicted in Fig. 17b. The figure implies a good agreement between the simulations and measurements results. The SIW cutoff frequency and null frequencies of $|S_{11}|$ are properly predicted. The measured insertion loss is less than 1.5 dB in the frequency range of 5.6-8.4 GHz, while in the same band, the measured return loss is better than 10 dB. The difference between $|S_{21}|_{meas}$ and $|S_{21}|_{sim}$ is less than 0.45dB.

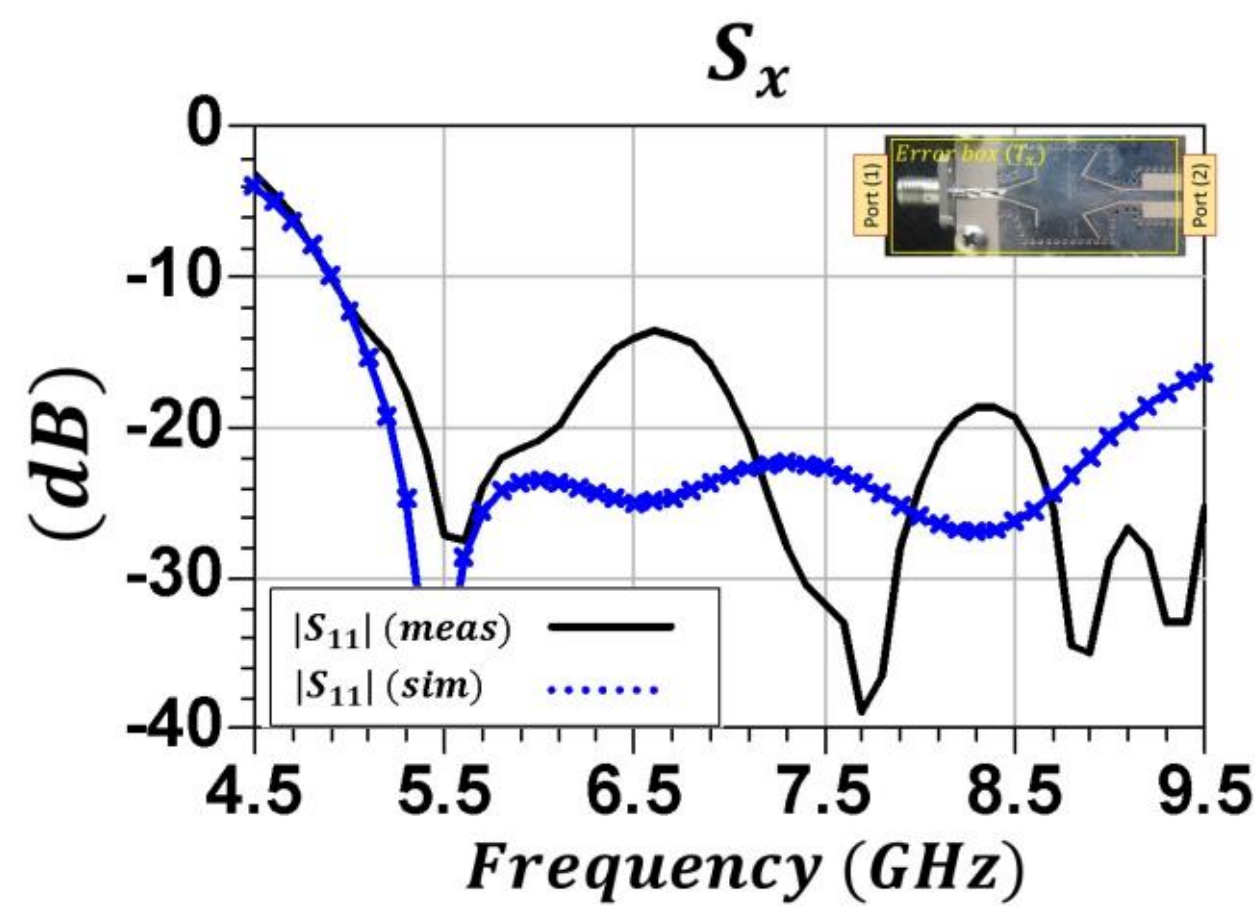

(a)

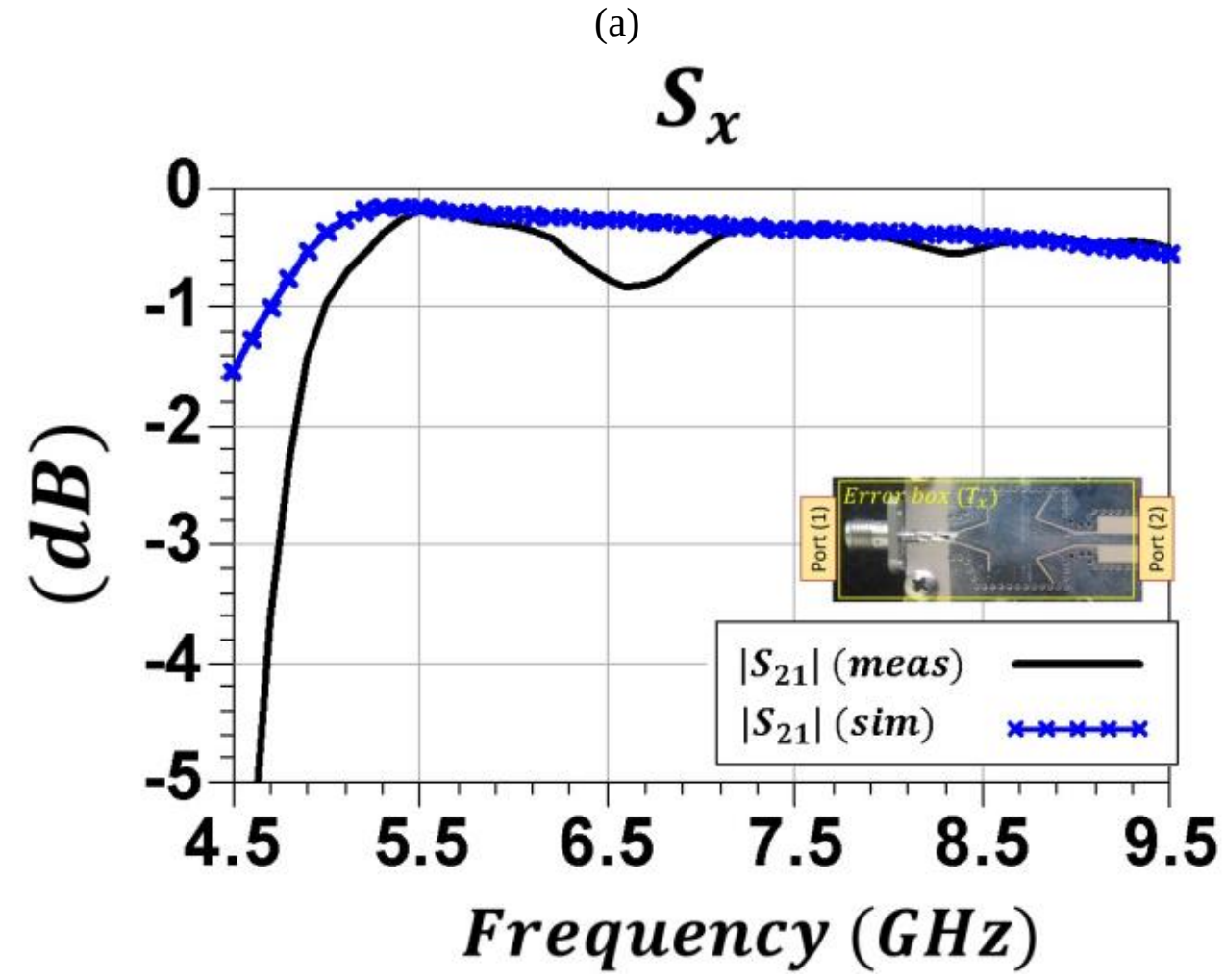

(b)

Fig. 18. S-parameters of the Error box: a) $|S_{11}|$, b) $|S_{21}|$..

Now we return to the calibration process. Since the depicted Error boxes of Fig. 17a are identical, the scattering transfer parameters of Through and Line circuits are

$$T_{Through} = T_x T_y \quad (14)$$

$$T_{Line} = T_x T_l T_y \quad (15)$$

respectively. We derive $T_{Through}$ and $T_{Line}$ from the measurement results of the calibration kit ($S_{Through}$ and $S_{Line}$). Now, if we assume a new matrix $T_{T.L}$ as

$$T_{T.L} = T_{Line}\left(T_{Through}\right)^{-1} = T_x T_l T_x^{-1} \quad (16)$$

it is evident from (16) that the eigenvalues of $T_{T.L}$ are the diagonal elements of $T_l$ while $T_x$ is a matrix whose columns are $T_{T.L}$ eigenvectors [26]. Figs. 18a, b illustrate the calculated S-parameters of the Error box ($T_x$). As the cutoff frequency of second guiding mode for the SIW is around 9.4 GHz, above this frequency, the excitation of the SIW second mode invalidates the calibration method. So, the figures are plotted in the frequency region of 4.5-9.5 GHz. It is seen from Fig. 18a, b that the return loss for the fabricated structure is better than 10 dB in the entire frequency band of 5-11 GHz and better than 15 dB in 5.2-9.4 GHz frequency range. The difference between insertion loss of the fabricated error box and that of simulated is less than 0.5 dB in the frequency region of 5.2-9.4 GHz. Corresponding to the increase in $|S_{11}|$ to about -15 dB, the insertion loss increases to about 0.8 dB around 6.5 GHz. It seems that the fabrication tolerances (specifically within the tapered slot transition area) are responsible for this little difference between the simulation and measurement results.

In the next step, if we call the scattering transfer matrix of the circuit of Fig. 16a $T_{Test}$, we can easily find that

$$T_{Test} = T_x T_{blocker} T_y \quad (17)$$

Since $T_y$ and $T_x$ networks are identical but mirrored, we have

$$T_y = (E_2 T_x E_2)^{-1} \quad (18)$$

where

$$E_2 = \begin{bmatrix} 0 & 1 \\ 1 & 0 \end{bmatrix} \quad (19)$$

So, the scattering transfer matrix of the transition is calculated as

$$T_{blocker} = T_x^{-1} T_{Test} E_2 T_x E_2 \quad (20)$$

The corresponding curves ($S_{blocker}$) are depicted in Figs. 19a, b. The overall behaviors of simulation and measurement S-parameters are similar. As seen, $|S_{11}|$ is less than -10 dB in the whole region of 5.6-8.4 GHz (40% of equivalent relative bandwidth). Also, the insertion loss is better than 1.5 dB in the same frequency band and better than 1 dB in 5.9-8.4 GHz (about 1/3 of relative bandwidth). The difference between $|S_{21}|$ of simulated and that of measured is less than 0.5 dB in 5.6-8.4 GHz. The small increase in the difference of the simulated and measured insertion losses are due to the fact that the attenuation constant which is derived from the applied calibration method is extremely sensitive to the precision of the measurement process. So, even very small errors in the measured values have noticeable effects on the calculated attenuation constant. Actually, for this error to be significantly suppressed, we need to use several different length Line circuits and implement the procedure presented in [30] which is out of the scope of this paper. Summarily, the fabricated DC-blocker presents an acceptable performance in 5.6-8.2 GHz which is approximately a 1/3 octave frequency region. Since according to Fig. 9c, its total length is only 5 mm or approximately $\frac{\lambda_g}{5}$ @ 7 GHz ($\lambda_g \approx 2.5\ cm$ @ 7 GHz for the SBCSL differential mode) this component can be easily used where a compressed phase-inverted differential mode DC-blocker is required. A short comparison of the proposed component to some other DC blocking solutions is presented in Table. VI. It is seen that the phase inversion functionality is only present in this work. Also, none of the other solutions have been designed for balanced TLs. Except for the ATC capacitor, the smallest feature size of our DC-blocker is much bigger than those of other solutions. In terms of the operational bandwidth, the ATC capacitor and the quarter-wave coupled-line have the best performance, but in the expense of cost and requirement of separate mounting phase in the former and very narrow gap size and feature size in the latter. In addition, the overall dimensions of the presented solutions are approximately in a same order of magnitude. Finally, the ATC capacitor has the best performance in terms of its insertion loss.

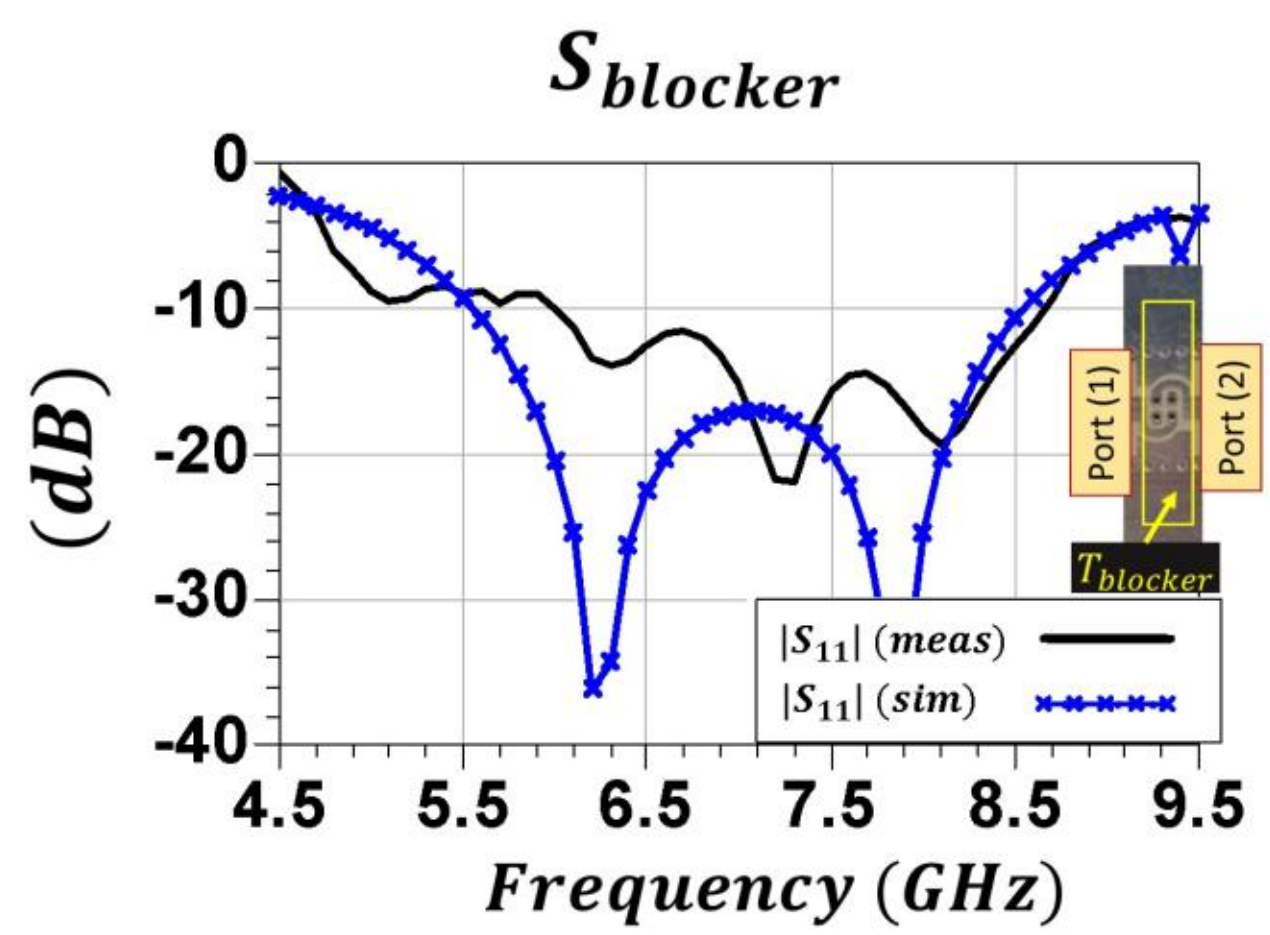

(a)

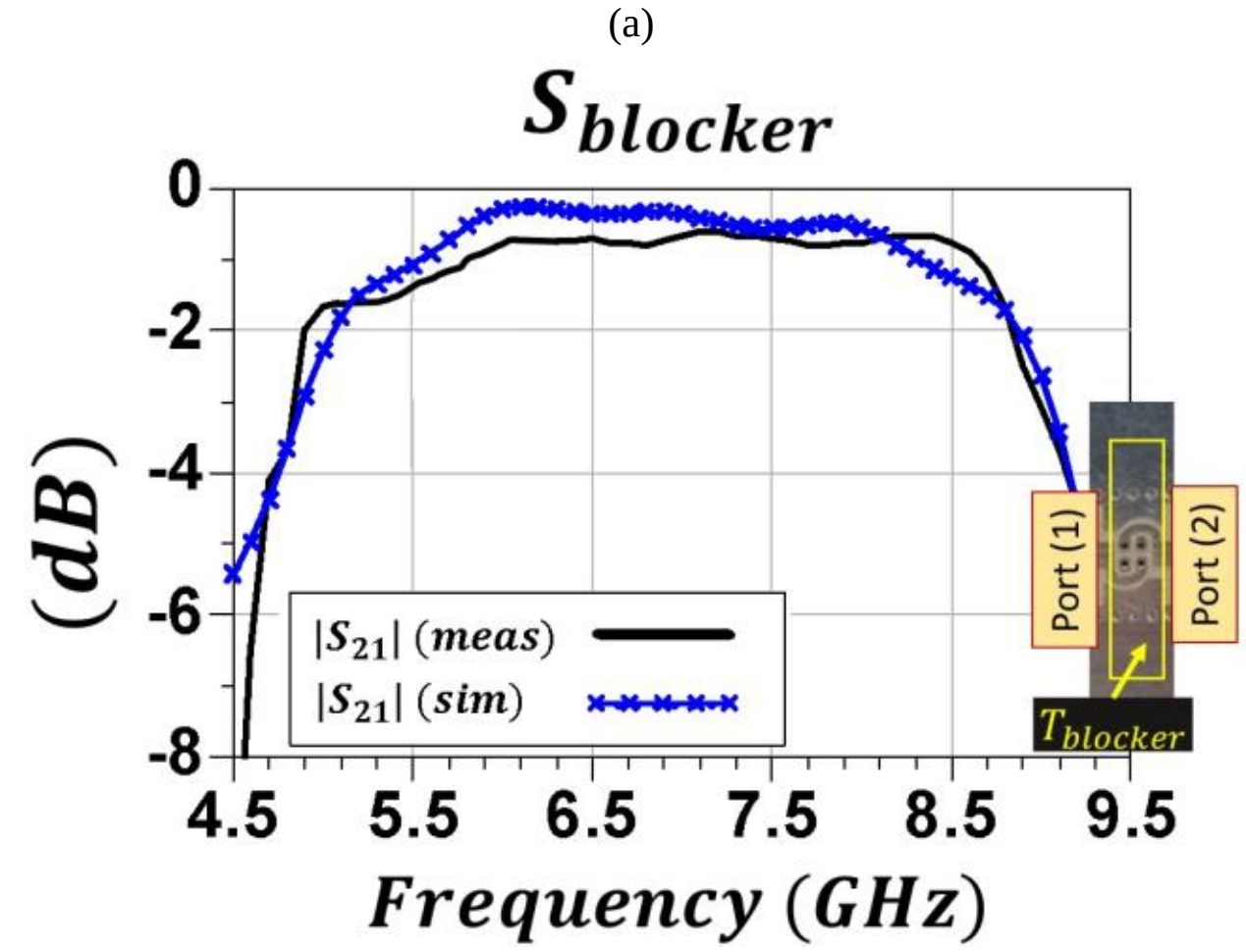

(b)

Fig. 19. S-parameters of the DC-blocker: a) $|S_{11}|$, b) $|S_{21}|$.

TABLE VI
COMPARISON OF OUR PROPOSED DC-BLOCKER WITH SOME OTHER AVAILABLE SOLUTIONS.

| DC-blocker type | Dimensions | Smallest feature size | 10 dB return loss bandwidth | Insertion loss | Single-phase fabrication | Balanced configuration | Phase inversion |
|---|---|---|---|---|---|---|---|
| Quarter-wave coupled-line [23] | $\frac{\lambda_g}{4} \times \frac{\lambda_g}{16}$ | $\frac{\lambda_g}{400}$ | 110% | 0.5 dB | yes | no | no |
| Interdigital capacitor [24] | $\frac{\lambda_g}{3.5} \times \frac{\lambda_g}{12}$ | $\frac{\lambda_g}{200}$ | 33% | 0.5 dB | yes | no | no |
| ATC 800b capacitor [25] | $\frac{\lambda_g}{9} \times \frac{\lambda_g}{9}$ | - | 100% | 0.4 dB | no | no | no |
| This work | $\frac{\lambda_g}{5} \times \frac{\lambda_g}{4}$ | $\frac{\lambda_g}{70}$ | 35% | 1 dB | yes | yes | yes |

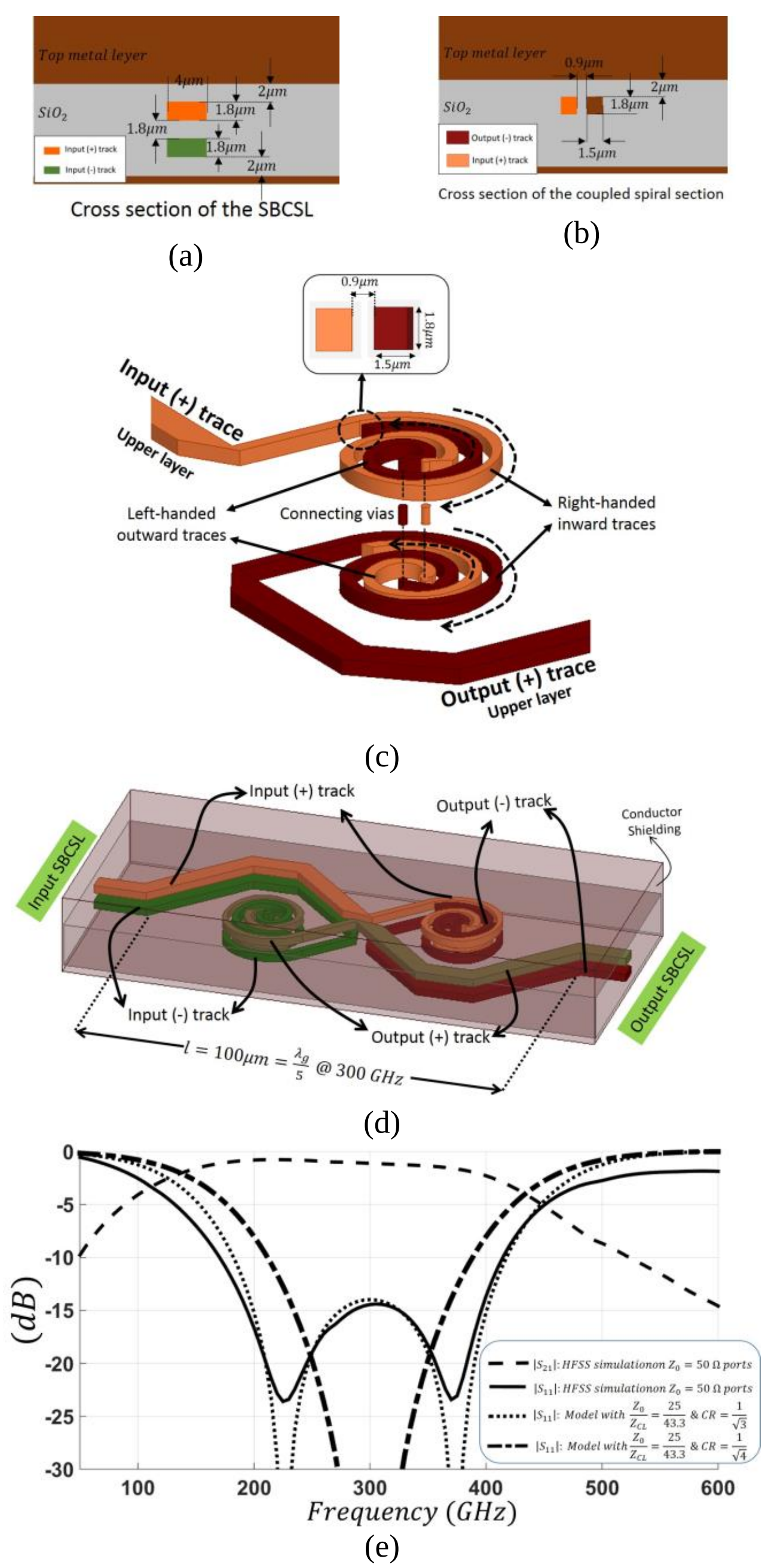

Fig. 20. The millimeter-wave IC DC-blocker: a) cross section of the SBCSL input TL completer structure, b) cross section of the coupled-line section c) layer-by-layer illustration of the coupled-spiral sections, d) paths of all four conductor traces of the input and output transmission lines, e) mathematical model response and Ansoft HFSS simulation results for part (d)

## V. Future Work

We can redesign our fabricated microwave DC-blocker for multi hundreds gigahertz IC applications. In this case both the guided wavelength and the circuit dimensions shrink by approximately comparable scales. Figs. 20a, b show the cross section of a coupled-line section suitable for CMOS IC DC-blocker realization, and a typical 50-ohm SBCSL transmission line, respectively. Both structures are realizable in an identical CMOS IC. The differential mode guided wavelength in the depicted SBCSL is around 500 $\mu m$ in 300 GHz which is more than one hundred times greater than the depicted traces widths and thicknesses and inter-trace spaces. So, the millimeter band IC DC-blocker will occupy very large area with respect to the SBCSL cross section and needs to be significantly miniaturized. To decrease the DC-blocker size without degrading its performance, we use the idea of spiral-form coupled structures which has been previously reported by authors [31], [32]. First, we stretch the input SBCSL (+) trace in an inward right-hand spiral-form in the upper layer. We put a metalized via at the spiral end and stretch the trace on the lower metal layer in an outward left-handed spiral-form. We set the total size of the open-ended trace at $\frac{\lambda_g}{4}$ at 300 GHz. This trace forms a coupled-line section if is put near another trace stretched right-handed inward in the lower metal layer and left-handed outward in the upper layer (Fig. 20c). To form the phase inverted balanced DC-blocker it is enough to place a similar spiral-form coupled-line structure in the input SBCSL (-) trace path, as depicted in Fig. 20d. The figure shows that the total length of the millimeter-wave IC DC-blocker is $\frac{\lambda_g}{5}$ at the center frequency which is similar to our fabricated microwave PCB DC-blocker. We simulate the structure in Ansoft HFSS and calculate $S_{11}$ at the shown input SBCSL differential port. The result that is shown in Fig. 20e illustrates that the input return loss is better than 10 dB in a170-420 GHz frequency band. Two more simulation results are also depicted in the figure which belong to DC-blockers comprised of coupled-line sections with $Z_{CL} \approx 43.3\ \Omega$ and coupling ratios of $CR = 0.5$ and $CR = \frac{1}{\sqrt{3}}$, respectively. Although the coupled-line sections of our spiral-form DC-blocker have coupling ratios of $CR = 0.5$, the overall spiral DC-blocker scattering parameters are similar to those of s simple balanced DC-blocker with $CR = \frac{1}{\sqrt{3}}$. It shows that the overall coupling ratio has been increased in our structure by using the proposed spiral configuration. It means that the space between coupled traces can be increased. As a consequence, the tolerance of the DC voltage difference will be significantly improved. Since more detailed discussion needs precise measurement results, we leave it hear for a possible future fabrication and measurement processes.

## VI. Conclusion

In this paper we first proposed a wave transfer matrix representation for a coupled-line configuration. We expressed the superiority of this method over other solutions for cascade circuits architecture. Then we theoretically describe that the couple-line structure is very suitable for the realization of DC-blockers inside balanced transition lines. We proposed a compact coupled-line DC-blocker with built-in phase inversion capability for 5-10 GHz applications. Also, a very wideband SIW-based coax-to-SBCSL balun transition was proposed which was successfully used for the measurement of the fabricated DC-blocker. Our measurement results show that both the DC-blocker and the balun transition can be used for differential signaling applications. In addition, we showed that the coupled-line phase inverted DC-blocker can be properly designed and integrated for millimeter-wave IC applications using a two-step spiral-form configuration.

## Acknowledgment

The authors would like to thank Iran National Scientific Foundation (INSF) for financial supporting this research work. The authors also would like to thank Mr. Ahmadi Ali-Abad of Fara-Afrand co. for his support in circuits fabrication.

## Appendix

The wave scattering transfer matrix of an infinitesimal length of a coupled-line TEM structure can be simply derived from its relevant equivalent circuit (Fig. A1a). Based on the figure notations, the voltages and currents have the following relations:

$$\begin{pmatrix} V_1 \\ V_2 \end{pmatrix} = -j\omega[L_e]\begin{pmatrix} I_3 \\ I_4 \end{pmatrix} + (I - \omega^2[L_e][C_e])\begin{pmatrix} V_3 \\ V_4 \end{pmatrix} \quad \text{(A1)}$$

$$\begin{pmatrix} I_1 \\ I_2 \end{pmatrix} = -\begin{pmatrix} I_3 \\ I_4 \end{pmatrix} + j\omega[C_e]\begin{pmatrix} V_3 \\ V_4 \end{pmatrix} \quad \text{(A2)}$$

where

$$[L_e] = \begin{bmatrix} L_1 & M \\ M & L_2 \end{bmatrix} \quad \text{(A3)}$$

$$[C_e] = \begin{bmatrix} C_1 + C_M & -C_M \\ -C_M & C_2 + C_M \end{bmatrix} \quad \text{(A4)}$$

The input and output power wave vectors can be derived from the corresponding voltages and currents defined at a real reference impedance $Z_0$, as

$$\begin{pmatrix} a_i \\ a_j \end{pmatrix} = \frac{1}{2\sqrt{Z_0}}\left(\begin{pmatrix} V_i \\ V_j \end{pmatrix} + Z_0\begin{pmatrix} I_i \\ I_j \end{pmatrix}\right) \quad \begin{cases} (i.j) \equiv (1.2) \\ \quad or \\ (i.j) \equiv (3.4) \end{cases} \quad \text{(A5)}$$

$$\begin{pmatrix} b_i \\ b_j \end{pmatrix} = \frac{1}{2\sqrt{Z_0}}\left(\begin{pmatrix} V_i \\ V_j \end{pmatrix} - Z_0\begin{pmatrix} I_i \\ I_j \end{pmatrix}\right) \quad \begin{cases} (i.j) \equiv (1.2) \\ \quad or \\ (i.j) \equiv (3.4) \end{cases} \quad \text{(A6)}$$

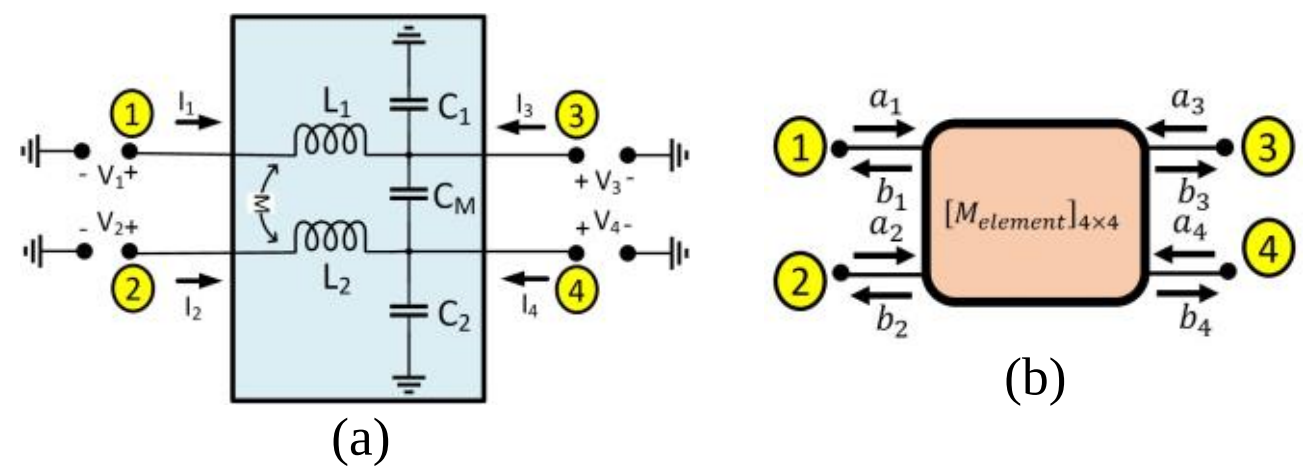

Fig. A1. The equivalent circuit of an infinitesimal length of a TEM coupled-line structure: (a) LC representation, (b) four-port equivalent network.

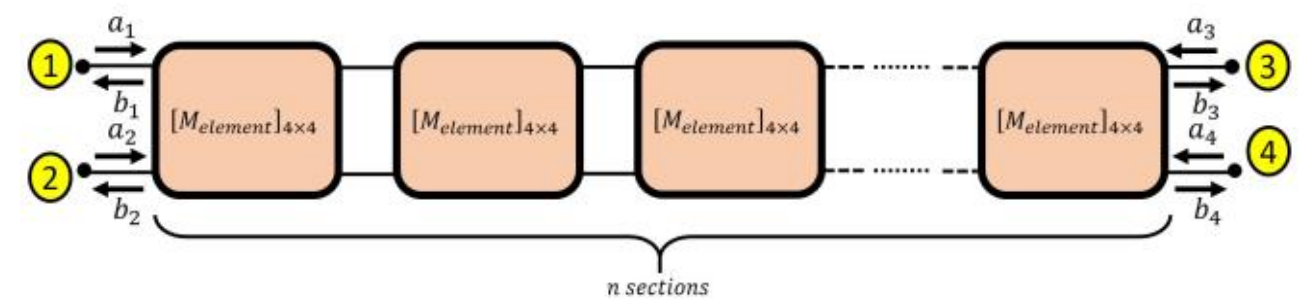

Fig. A2. The equivalent circuit of an infinitesimal length of a TEM coupled-line structure: (a) LC representation, (b) four-port equivalent network.

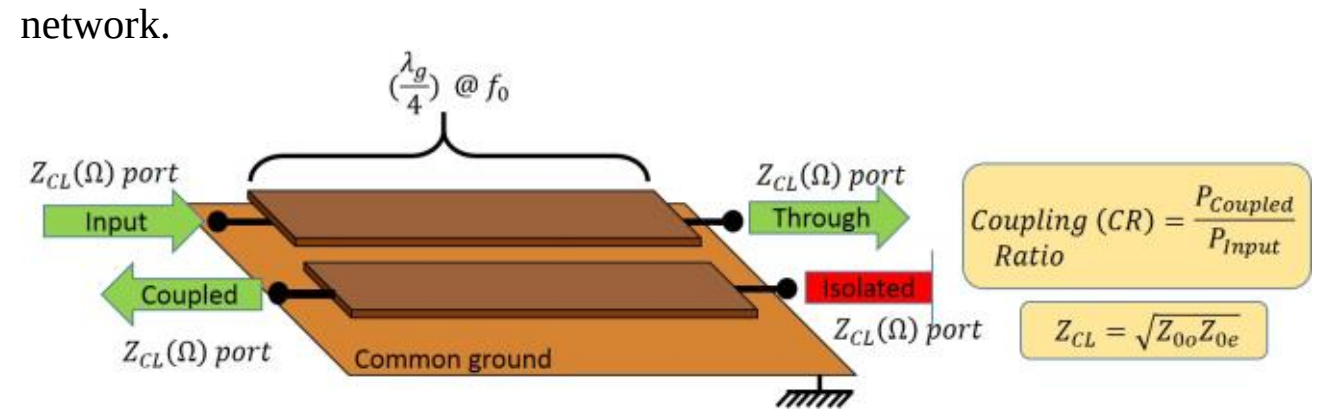

Fig. A3. A pair of quarter-wavelength couple-line as a directional coupler.

We consider the circuit of Fig. A1a as a four-port network of Fig. A1b and assign a four-dimensional wave scattering transfer matrix ($M_{element}$) to it, as

$$\begin{pmatrix} a_1 \\ a_2 \\ b_1 \\ b_2 \end{pmatrix} = M_{element}\begin{pmatrix} b_3 \\ b_4 \\ a_3 \\ a_4 \end{pmatrix} \quad \text{(A7)}$$

Eq. (A1) to (A7) result in the following relation for $M_{element}$

$$\begin{aligned} M_{element} = I_{4\times4} &- \frac{\omega^2}{2}\begin{bmatrix} I_{2\times2} \\ I_{2\times2} \end{bmatrix}[L_e][C_e][I_{2\times2} \quad I_{2\times2}] \\ &+ \frac{j\omega}{2}\left(Z_0\begin{bmatrix} I_{2\times2} \\ -I_{2\times2} \end{bmatrix}[C_e][I_{2\times2} \quad I_{2\times2}] \right. \\ &\left. + \frac{1}{Z_0}\begin{bmatrix} I_{2\times2} \\ I_{2\times2} \end{bmatrix}[L_e][I_{2\times2} \quad -I_{2\times2}]\right) \end{aligned} \quad \text{(A8)}$$

In (A8) $I_{4\times4}$ represents a $(4\times4)$ identity matrix. (A8) can be simplified since we have assumed a lossless TEM wave transmission along the coupled-line structure. This assumption which is a good primary approximation for the case of PCB and IC transmission line structures has an important outcome: Both the scalar electric potential and longitudinal part of the vector magnetic potential satisfy Laplace's equation. It means that if we define a capacitance per unit length matrix ($[C_u]$) and an inductance per unit length matrix ($[L_u]$) for the coupled-line structure, these two matrices are related to each other by

$$[L_u] = \frac{1}{v_{TL}^2}[C_u]^{-1} \quad \text{(A9)}$$

where $v_{TL}$ is the velocity of electromagnetic wave propagation along our transmission line. These matrices are related to $[C_e]$ and $[L_e]$ by

$$[C_u] = \frac{[C_e]}{\left(\frac{l}{n}\right)} \quad \text{(A10)}$$

$$[L_u] = \frac{[L_e]}{\left(\frac{l}{n}\right)} \quad \text{(A11)}$$

where $l$ is the total length of the coupled-line section and $\frac{l}{n}$ is the infinitesimal length associated with the coupled-line element of Fig. A1a. Using (A9)-(A11) and by defining the electrical length parameter ($\theta$) as

$$\theta = \frac{\omega l}{v_{TL}} \quad \text{(A12)}$$

Eq. A8 can be rewritten as:

$$M_{element} = I_{4\times4} - \left(\frac{\theta}{n}\right)^2 [D^r] + j\left(\frac{\theta}{n}\right)[D^i] \quad \text{(A13)}$$

where

$$[D^r] = \left(\frac{1}{2}\right)\begin{bmatrix} I_{2\times2} & I_{2\times2} \\ I_{2\times2} & I_{2\times2} \end{bmatrix} \quad \text{(A14)}$$

$$[D^i] = \left(\frac{1}{2}\right)\left(\begin{bmatrix} I_{2\times2} \\ -I_{2\times2} \end{bmatrix}(v_{TL}Z_0[C_u])[I_{2\times2} \quad I_{2\times2}] + \begin{bmatrix} I_{2\times2} \\ I_{2\times2} \end{bmatrix}(v_{TL}Z_0[C_u])^{-1}[I_{2\times2} \quad -I_{2\times2}]\right) \quad \text{(A15)}$$

$[D^r]$ and $[D^i]$ matrices contains information of power distribution inside the four-port network. By this definition, the wave scattering transfer matrix of a cascade connection of $n$ four-port networks (like that of Fig. A1b) in the style depicted in Fig. A2 is simply equal to:

$$M_{Total} = M_{element}^n = \left(I_{4\times4} - \left(\frac{\theta}{n}\right)^2 [D^r] + j\left(\frac{\theta}{n}\right)[D^i]\right)^n \quad \text{(A16)}$$

if $n \to \infty$ the contribution of $[D^r]$ vanishes and

$$\lim_{n\to\infty} M_{Total} = \lim_{n\to\infty}\left(I_{4\times4} + j\left(\frac{\theta}{n}\right)[D^i]\right)^n = e^{j\theta[D^i]} \quad \text{(A17)}$$

This shows that the scattering transfer matrix of a coupled-line section can be described in the form of a matrix exponential. Since $[D^i]$ is an involutory matrix, we have

$$[D^i]^2 = I_{4\times4} \quad \text{(A18)}$$

This characteristic is useful in Taylor series representation of $M_{Total}$:

$$M_{Total} = e^{j\theta[D^i]} = \cos(\theta[D^i]) + j * \sin(\theta[D^i]) = \left([I]_{4\times4} - \frac{\theta^2[D^i]^2}{2!} + \frac{\theta^4[D^i]^4}{4!} - \frac{\theta^6[D^i]^6}{6!} + \cdots\right) + j\left(\theta[D^i] - \frac{\theta^3[D^i]^3}{3!} + \frac{\theta^5[D^i]^5}{5!} - \frac{\theta^7[D^i]^7}{7!} + \cdots\right) \quad \text{(A19)}$$

or

$$M_{Total} = \cos(\theta) * I_{4\times4} + j\sin(\theta) * [D^i] \quad \text{(A20)}$$

We can look at the coupled-line section as a $Z_{CL}$-matched coupled-line structure (Fig. A3). If we suppose that the coupled-line structures are symmetric ($[C_u]$ becomes a symmetrical matrix) we have [16]:

$$Z_{0o} = \frac{\left(\frac{l}{n}\right)}{v_{TL}(C_1 + 2C_M)} = Z_{CL}\sqrt{\frac{1-CR}{1+CR}} \quad \text{(A21)}$$

$$Z_{0e} = \frac{\left(\frac{l}{n}\right)}{v_{TL}C_1} = Z_{CL}\sqrt{\frac{1+CR}{1-CR}} \quad \text{(A22)}$$

where $Z_{0o}$ and $Z_{0e}$ are odd and even impedances of the coupled-line structure and $C_1$ and $C_M$ are related to $[C_u]$ by (A4) and (A10) and CR is the coupling ratio of the quarter-wavelength couple-line structure if used as a directional coupler in the configuration of Fig. A3. By some mathematical manipulation, we obtain:

$$[D^i] = \frac{1}{2\sqrt{1-CR^2}}\left(\left(\frac{Z_0}{Z_{CL}}\right)\begin{bmatrix} I_{2\times2} \\ -I_{2\times2} \end{bmatrix}\begin{bmatrix} 1 & -CR \\ -CR & 1 \end{bmatrix}[I_{2\times2} \quad I_{2\times2}] + \left(\frac{Z_{CL}}{Z_0}\right)\begin{bmatrix} I_{2\times2} \\ I_{2\times2} \end{bmatrix}\begin{bmatrix} 1 & CR \\ CR & 1 \end{bmatrix}[I_{2\times2} \quad -I_{2\times2}]\right) \quad \text{(A23)}$$

By this interpretation $M_{Total}$ can been directly expressed in terms of $\theta$, $CR$ and $\frac{Z_0}{Z_{CL}}$ using (A20).

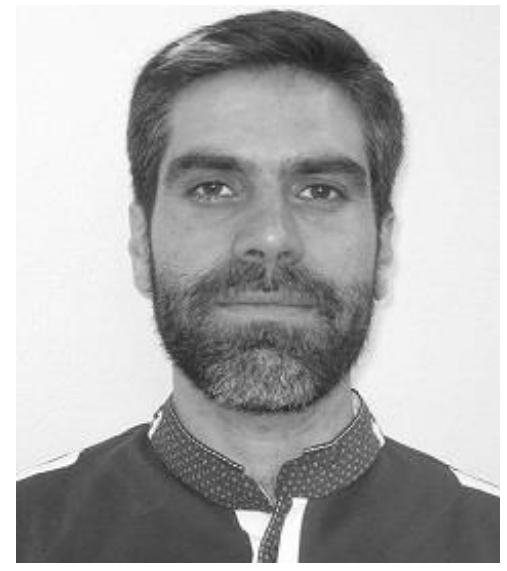

**Mostafa Abdolhamidi** was born in Tehran, Iran. He received the B.S. and M.S. degrees in Fields and Waves in Communication engineering from University of Tehran, Iran in 2004 and 2008 respectively, and now is working toward Ph.D. in the same field and the same university. From 2008 he has started a close collaboration with Fara-Afrand Co. in Tehran where he has focused on the design and manufacture of different passive and active RF circuits for high power terrestrial video and audio broadcasting applications. During these years the company was successful in the production of the first multi-kilowatt national digital video broadcast transmitter system and has continued to the mass production of nationwide high power broadcasting systems and sub-systems. In parallel to the industrial activity, Mr. Abdolhamidi started to work on leading-edge millimeter-wave technology during his Ph.D. where he is designing novel passive interconnects for IC millimeter-wave applications.

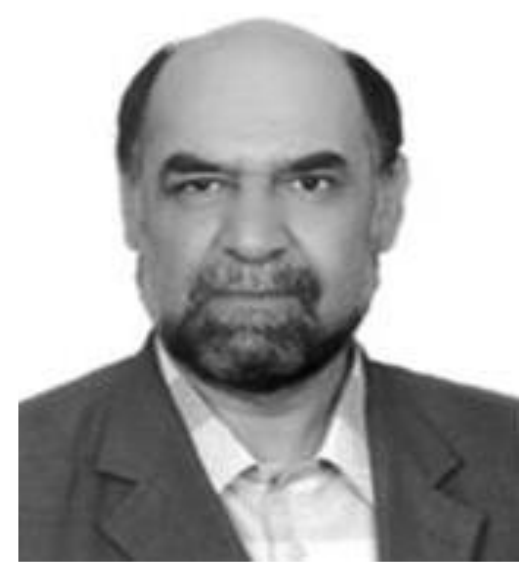

**Mahmoud Mohammad-Taheri** was born in Saryazd, Yazd, Iran. He received the B.Sc. degree in electrical engineering from the Sharif University of Technology, Tehran, Iran, in 1979, and the M.Sc. degree in telecommunication systems and the Ph.D. degree in microwave engineering from the University of Essex, Essex, U.K., in 1986 and 1990, respectively. From 1982 to 1985 and 1990 to 1991, he was with the Telecommunication Research Center as a Microwave Research Engineer and Project Manager. From 1997 to 2001, he was involved with the Oil Ministry of Iran as a consultant in offshore and onshore telecommunication projects. From October 2001 to November 2002, he was on sabbatical leave with the University of Waterloo, where he taught electromagnetic and he was involved with the design and analysis of ultra-wideband distributed amplifiers (DAs). From August 2008 to September 2011, he was a Post-Doctoral Fellow with the University of Waterloo, worked in the area of design and fabrication of millimeter wave low noise amplifier. Since 1991, he has been a member of the School of Electrical and Computer Engineering, Faculty of Engineering, University of Tehran, Tehran, Iran, where he is currently an Associate Professor. He has authored or co-authored over 60 papers in well-known journals and conferences. He has authored four books. His research interests are multimode dielectric-loaded cavity microwave filters, radio wave propagation, microwave remote sensing, ultra-wideband DAs, and millimeter and sub-millimeter wave low noise amplifiers.